\documentclass{article}

    \PassOptionsToPackage{numbers, compress}{natbib}

    \usepackage[final]{neurips_2025}

\usepackage[utf8]{inputenc} 
\usepackage[T1]{fontenc}    
\usepackage{hyperref}       
\usepackage{url}            
\usepackage{booktabs}       
\usepackage{amsfonts}       
\usepackage{nicefrac}       
\usepackage{microtype}      
\usepackage{xcolor}         

\usepackage{sidecap}
\usepackage{xcolor}
\usepackage{microtype}
\usepackage{graphicx}
\usepackage{subfigure}
\usepackage{booktabs} 
\usepackage{hyperref}
\usepackage{url}
\usepackage{xurl}
\usepackage{svg}
\usepackage[most]{tcolorbox} 
\usepackage[ruled,vlined]{algorithm2e}
\usepackage{multicol}
\usepackage{wrapfig,lipsum,booktabs}
\usepackage[capbesideposition=outside,capbesidesep=quad]{floatrow}
\hypersetup{
           breaklinks=true,   
           colorlinks=true,   
        }

\usepackage{amsmath}
\usepackage{amssymb}
\usepackage{mathtools}
\usepackage{amsthm}
\usepackage{float}
\usepackage{enumitem} 
\usepackage{listings}
\usepackage{uarial}
\tcbset{
  mybox/.style={
    colframe=blue!40!black,     
    breakable,                  
    enhanced,                   
  }
}

\title{AgentBreeder: Mitigating the AI Safety Risks of Multi-Agent Scaffolds via Self-Improvement}

\author{J Rosser\\
University of Oxford\\
\texttt{jrosser@robots.ox.ac.uk}
 \\
\And
Jakob Foerster \\
FLAIR \\
University of Oxford \\
}

\begin{document}

\maketitle

\begin{abstract}
  Scaffolding Large Language Models (LLMs) into multi-agent systems often improves performance on complex tasks, but the safety impact of such scaffolds has not been thoroughly explored. We introduce \textsc{AgentBreeder}, a framework for multi-objective self-improving evolutionary search over scaffolds. We evaluate discovered scaffolds on widely recognized reasoning, mathematics, and safety benchmarks and compare them with popular baselines. In `blue' mode, we see a 79.4\% average uplift in safety benchmark performance while maintaining or improving capability scores. In `red' mode, we find adversarially weak scaffolds emerging concurrently with capability optimization. Our work demonstrates the risks of multi-agent scaffolding and provides a framework for mitigating them. Code is available at \url{https://github.com/jrosseruk/AgentBreeder}.
\end{abstract}

\section{Introduction}
\label{introduction}

Recently, the field of artificial intelligence has witnessed remarkable advancements in Large Language Models (LLMs) and their applications \citep{zhao2023survey}. LLMs are capable of exhibiting human-like reasoning \citep{amirizaniani2024can, sun2024determlr, towardslargereasoningmodels}, enabling their application beyond natural language processing to diverse areas such as code generation \citep{funsearch, wang2023review, yeticstiren2023evaluating}, embodied AI in robotics \citep{hu2023toward, kong2024superalignment, sartor2024neural}, and autonomous agents \citep{operator, proxy}. Our research is motivated by accelerated advancements in autonomous agents such as the recent release of Operator \citep{operator} and Proxy \citep{proxy} - agents that browse the web and perform tasks autonomously on behalf of the user. Alignment research to date has almost exclusively focused on the safety of LLMs in unipolar scenarios; ensuring a single LLM remains aligned inside a single-agent system. When deployed on the web, agents are placed in novel multi-agent scaffolds and subjected to multi-polar challenges  \citep{multipolar}. With highly-capable agents now being deployed at scale, we seek to address the immediate need for more comprehensive safety evaluations of multi-agent systems.

In this paper, we introduce \textsc{AgentBreeder}, an evolutionary open-ended framework capable of generating large populations of diverse multi-agent scaffolds. By equipping this framework with multi-objective optimization, we explore the generation of multi-agent scaffolds along complementary objectives of capability and safety. \textsc{AgentBreeder} can be used to blue team a set of scaffolds to generate offspring that exhibit greater adversarial robustness and performance on capability benchmarks. Similarly, a red teaming approach generates offspring that exhibit greater vulnerability to adversarial attacks.

\begin{figure}[!ht]
\centering
\includegraphics[width=0.95\textwidth]{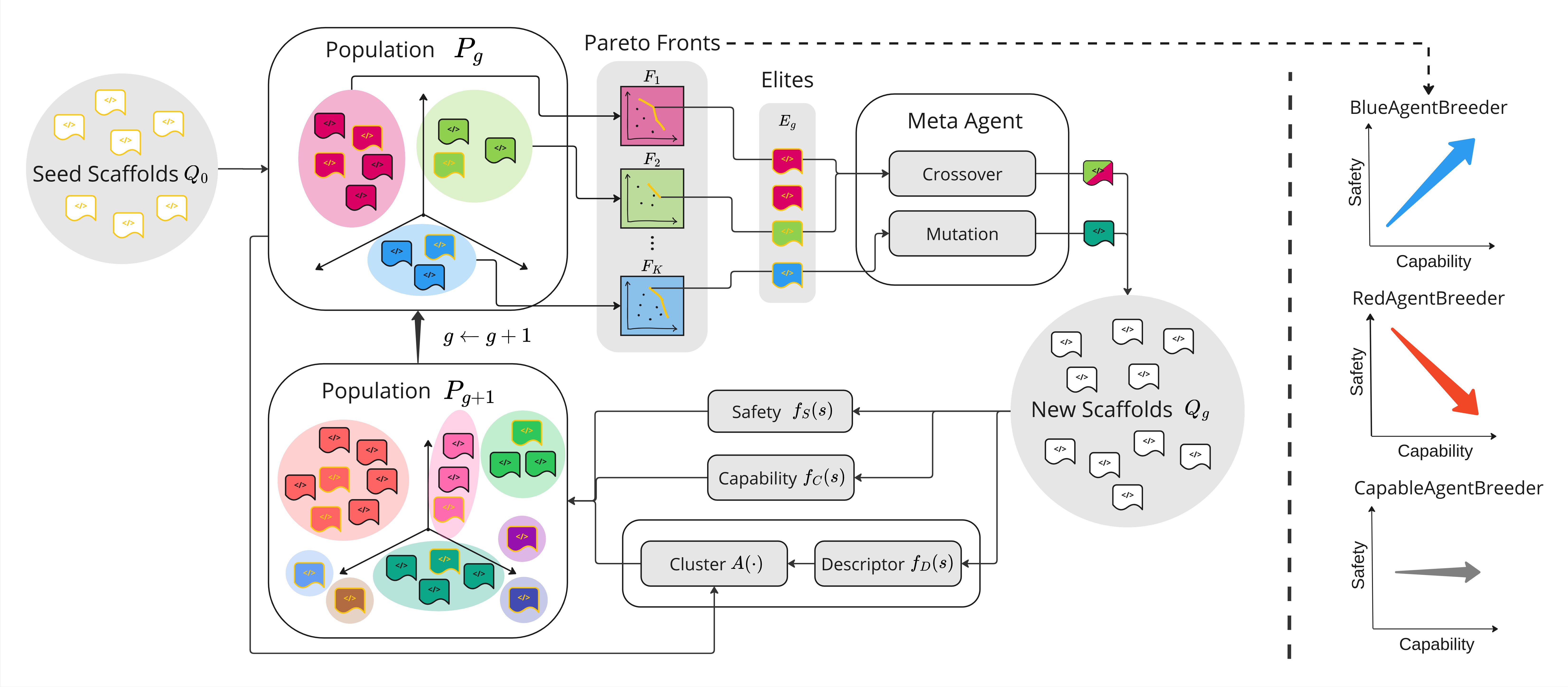}
\caption{A high‐level illustration of the \textsc{AgentBreeder} algorithm as outlined in Algorithm \ref{alg:agentbreeder_simplified}. Starting from seed scaffolds $Q_0$, at each generation $g$ the newly generated scaffolds $Q_{g-1}$ are evaluated on capability ($f_C(s)$) and/or safety ($f_S(s)$) benchmarks, then embedded via $f_D(s)$ for clustering $A(\cdot)$ into $K$ clusters. Within each cluster, Pareto fronts  ${F_1, ..., F_K}$ are identified according to $f_C(s)$ and/or $f_S(s)$, and these ``frontier'' solutions become the elite set $E_g$. An LLM-based Meta Agent applies crossover and mutation to the elites, creating new offspring scaffolds $Q_g$. These offspring are added to the population for the next generation $P_{g+1}$. By repeating this process for $G$ generations, \textsc{AgentBreeder} explores a large, diverse set of multi‐agent systems while balancing capability and safety. \textsc{AgentBreeder} can be run in 3 different modes and the right-hand side of this figure shows the optimal direction of travel of the Pareto front for each generation. \textsc{BlueAgentBreeder} is a defense mode and seeks to maximize both capability and safety, whereas \textsc{RedAgentBreeder} is an attack mode minimizing safety. \textsc{CapableAgentBreeder} serves as our baseline, only optimizing for capability without regard to safety.}
\label{fig: AgentBreeder}
\end{figure}
\begin{algorithm}[H]

\caption{AgentBreeder}
\label{alg:agentbreeder_simplified}

\textbf{Input:} Number of generations $G$; Number of clusters $K$; Number of evolutions $M$; Capability benchmark $f_{C}(s)$; Safety benchmark $f_{S}(s)$; Embedding function $f_{D}(\cdot)$; Seed scaffolds $Q_0$; Clustering function $A(\cdot)$.

\textbf{Initialize seed population $P_0$ = $Q_0$ of size $N_0$.}
\For{generation $g = 1$ to $G$}{

  \For{scaffold $s \in Q_{g-1}$}{
  \begin{enumerate}
    \item Compute capability $f_{\text{C}}(s)$ and safety $f_{\text{S}}(s)$.
    \item Compute embedding $e_s \leftarrow f_{D}(s)$.
  \end{enumerate}
  }

  \textbf{Cluster population into $K$ clusters:} $C_1, C_2, \ldots, C_K \leftarrow A(e_1, e_2, ..., e_{N_g})$.

  \textbf{Identify Pareto Elites $E_g$:}
     \begin{enumerate}
      \item Set $E_g \leftarrow \emptyset$.
      \item \For{cluster $k = 1$ to $K$}{
        \begin{enumerate}
            \item Find its Pareto front $F_k$ using $f_{\text{C}}$ and $f_{\text{S}}$.
            \item Update elite cohort $E_g \leftarrow E_g \cup F_k$.
        \end{enumerate}
      }
      \end{enumerate}
  
  \textbf{Generate offspring $Q_g$:}
  \begin{enumerate}
    \item Set $Q_g \leftarrow \emptyset$.
    \item \For{evolution $m = 1$ to $M$}{
      \begin{enumerate}
        \item Weighted sampling 1 or 2 elites from $E_g$.
        \item If 2 elites, Meta Agent performs \text{Crossover}; 
        otherwise \text{Mutation}.
        \item Add the offspring to $Q_g$.
      \end{enumerate}
    }
  \end{enumerate}

  \textbf{Update population:} $P_g \leftarrow P_{g-1} \cup Q_g$.
  \textbf{Update population size:} $N_g \leftarrow N_{g-1} + M$.
}
\textbf{Output:} Final population $P_G$.

\end{algorithm}

\textbf{Our main contributions are listed as follows:}

\begin{itemize}
    \item \textbf{Attack.} We introduce a novel red teaming method which can be used to explore the attack surfaces of base LLMs when deployed in multi-agent settings.
    \item \textbf{Defense.} We introduce a novel blue teaming method for generating multi-agent scaffolds that exhibit greater robustness to adversarial attacks.
    \item \textbf{Evaluation.} We implement \textsc{AgentBreeder} in Inspect \citep{AISIInspect} to ensure the reproducibility and extensibility of our results and methods. 
\end{itemize}

\section{Background}
\label{background}

\textbf{Multi-Agent Systems.} Multi-agent systems consist of multiple interacting intelligent agents such as LLM assistants like ChatGPT \citep{gpt4omini}. These systems offer several advantages over single-agent approaches \citep{yang2024multi}, including planning \citep{li2024agent, cao2024llm}, task decomposition \citep{chen2023agentverse, fourney2024magentic, ghafarollahi2024sciagents, qian2023communicative}, and specialization \citep{chan2023chateval, chen2023agentverse, fourney2024magentic, ghafarollahi2024sciagents, qian2023communicative}. The terms ``multi-agent system'', ``multi-agent framework'', ``agent'' and ``scaffold'' are used interchangeably in literature to refer to the structural frameworks that support communication between multiple LLMs \citep{maliciouserrormultiagentsystems, godel, adas, scaffoldedllms}. In this paper, we will primarily use the term ``scaffold'' to refer to the architectures - often defined in Python code - that support the operation of multi-agent systems. 

\textbf{Automated Design of Agentic Systems.} We build upon the seminal work of \citet{adas} which introduces the research area Automated Design of Agentic Systems (ADAS), an automated approach to discovering high-performing (multi-agent) scaffolds. \citet{adas} formulate ADAS as an optimization algorithm comprising 3 key components; the search space, the search algorithm and the evaluation function. \citet{adas} also propose a search algorithm called ``Meta Agent Search'' where a single ``Meta Agent'' discovers scaffolds by programming them in Python code. Python is a Turing Complete language \citep{Boyer1983TuringLISP} therefore searching within a code space allows the Meta Agent to program theoretically any possible scaffold. This approach has shown promising results \citep{adas, godel}, with discovered scaffolds outperforming state-of-the-art hand-designed baselines across various tasks, including reading comprehension, mathematics, and science questions \citep{mmlu, mgsm, gpqa, ARC-AGI, drop}.

We formulate \textsc{AgentBreeder} with respect to the ADAS methodology. We replicate the approach of \citet{adas} by seeding our population with hand-designed scaffolds. We prompt a single ``Meta Agent'' to search for novel scaffolds in the space of Python code. We introduce a novel quality-diversity search algorithm inspired by MAP-Elites \citep{mapelites}, where the Meta Agent evolves new scaffolds via the random sampling, mutation and crossover of the highest performing individual or ``elite'' of each niche of the population. We cluster scaffolds based on their architectural features, and evaluate the performance of scaffolds on two benchmarks, one for capability and one for safety. We employ multi-objective optimization, sampling elites from the Pareto front of each cluster.

\textbf{Multi-Objective Evolutionary Algorithms.} Multi-objective optimization searches for solutions to problems with multiple, often conflicting objectives. Multi-objective evolutionary algorithms (MOEAs) incorporate an evolutionary approach to generate a diverse set of solutions \citep{moea}. In \textsc{AgentBreeder} we seek to balance the objectives of capability and safety whilst evolving a diverse range of scaffolds. \textsc{AgentBreeder} balances quality and diversity by clustering scaffolds based on their architectural features and randomly sampling elites from each cluster's capability-safety Pareto front. A solution is Pareto optimal if no other solution improves one objective without worsening another. The Pareto front comprises all such optimal solutions.

\textbf{Adversarial Robustness.} Adversarial robustness quantifies the resilience of a model or scaffold to malicious inputs such as jailbreaks \citep{chao2023jailbreaking} and prompt injection \citep{injection}. Red teaming, the practice of simulating adversarial scenarios to identify vulnerabilities, has emerged as a crucial tool for assessing AI model risks and alignment \citep{rainbowteaming, perez2022red}. In \textsc{RedAgentBreeder}, instead of generating adversarial examples, we seek to evolve multi-agent scaffolds that are more vulnerable to adversarial attacks than the base model. In \textsc{BlueAgentBreeder}, we seek to evolve multi-agent scaffolds that are more robust to adversarial attacks than the base model.

\section{Related Work}
\label{relatedwork}

\textbf{Self-Referential Self-Improving Systems.} Numerous frameworks \citep{yuan2024evoagent, evomac, adas, comfybench, godel} have been proposed to address the design of multi-agent scaffolding. EvoAgent \citep{yuan2024evoagent} extends single expert agents to multi-agent scaffolds via evolutionary algorithms, whilst \textsc{AgentBreeder} evolves the entire system as a unit. EvoMAC \citep{evomac} evolves agents and their connections during test time to improve code generation, whereas \textsc{AgentBreeder} is domain agnostic and can explore the entire search space of scaffolds. ADAS \citep{adas}, ComfyAgent \citep{comfybench} and Gödel Agent \citep{godel} search in the space of code for novel scaffolds, but unlike \textsc{AgentBreeder} they do not incorporate a quality-diversity mechanism for exploring agent design space. FunSearch \citep{funsearch} is an evolutionary method to search the function space for high-performing computer programs but not necessarily scaffolds. PromptBreeder \citep{promptbreeder} is an evolutionary self-improving framework that evolves prompts for a given domain, but does not focus on the scaffold as a whole.

\textbf{Multi-Agent Safety Research.} \citet{psysafe} evaluate the safety of multi-agent scaffolds from a psychological perspective by injecting agents with malicious traits, and provide mitigation strategies such as performing psychological assessments and therapy for agents. Polaris \citep{polaris} introduces a safety-focused scaffold for real-time patient healthcare conversations.
\citet{maliciouserrormultiagentsystems} explore the resilience of multi-agent scaffolds when injected with malicious or error-prone agents. \citet{scaffoldedllms} provide a more thorough discussion of the safety risks associated with scaffolded LLMs.

\section{AgentBreeder}
\label{agentbreeder}

We now introduce \textsc{AgentBreeder}, our automated, evolutionary approach to discovering new multi-agent scaffolds. By evolving a large, diverse corpus of multi-agent scaffolds, \textsc{AgentBreeder} seeks to address the challenge of evaluating the vulnerabilities of base LLMs acting inside capability-optimized multi-agent scaffolds. The pseudo-algorithm is given in Algorithm \ref{alg:agentbreeder_simplified} and Figure \ref{fig: AgentBreeder} provides a brief overview. \textsc{AgentBreeder} can be run in three modes:
\begin{itemize}
    \item \textsc{BlueAgentBreeder} - In this mode, the Meta Agent adopts the role of a ``Blue Team", searching for scaffolds that exhibit high capability and safety when evaluated on representative benchmarks. 
    \item \textsc{RedAgentBreeder} - In this mode, the Meta Agent adopts the role of a ``Red Team", minimizing performance on one safety benchmark whilst maximizing performance on one capability benchmark.
    \item \textsc{CapableAgentBreeder} - In this mode, the Meta Agent seeks to maximize performance on a single capability benchmark without consideration of safety.
\end{itemize}

\subsection{Seed Scaffolds}
Following the approach of \citet{adas} and \citet{godel}, we seed our population with the same 7 hand-designed scaffolds. These comprise Chain-of-Thought (CoT) \citep{cot}, Self-Consistency with Chain-of-Thought \citep{selfconsistencycot}, Self-Refine \citep{selfrefine}, LLM-Debate \citep{debate}, Step-back Abstraction \citep{stepback}, Quality-Diversity (QD) \citep{aiscientist}, and Role Assignment \citep{roleassignment}. Before running our evolution on our chosen benchmark, we evaluate a single CoT agent on 1,000 samples from the validation set of the benchmark, oversampling and resampling where necessary. For each generation, we validate the newly discovered scaffolds using a balanced sampling strategy, selecting 50\% positive and 50\% negative samples. Positive samples correspond to those the baseline CoT agent answered correctly and vice versa. We implemented this balanced sampling strategy specifically to increase the signal strength for the evolutionary process, ensuring adequate representation of both success and failure cases. Often improvements between generations are marginal, so this method increases information gain by providing a stronger signal for the evolutionary process.

\subsection{Mutation Operators}

\textsc{AgentBreeder}'s evolutionary search algorithm mimics the process of natural selection comprising mutation, crossover and selection. Claude 3.5 Sonnet \citep{claude} (\textit{claude-3-5-sonnet-20241022-v2:0}) is used as the core model of the Meta Agent due to its state-of-the-art performance on code generation tasks \citep{scicode}.

\textbf{Selection.} Selection pressure is applied at each generation by sampling candidate scaffolds at random from the Pareto fronts of each cluster. In \textsc{CapableAgentBreeder}, the Pareto front is simply the elite of each cluster, whereas in \textsc{BlueAgentBreeder} and \textsc{RedAgentBreeder}, the Pareto front comprises the scaffolds which best trade-off safety and capability.

\textbf{Mutation.} The Meta Agent uses weighted random sampling to select either the crossover or mutation operator. Weighting the mutation operator twice as highly as crossover was found empirically to lead to faster convergence. Mutation is performed via random sampling of mutation operators expressed as short textual passages we hand-designed. There are two types of mutation operators, capability-enhanced and safety-enhanced. When running \textsc{BlueAgentBreeder}, mutation operators are randomly sampled from the concatenated capability- and safety-enhanced corpus. In \textsc{RedAgentBreeder} and \textsc{CapableAgentBreeder} the safety-enhanced operators are omitted.  The full list of Meta Agent prompts and mutation operators are given in Appendix \ref{prompts}.

\textbf{Crossover.} During crossover, the Meta Agent is provided with two randomly sampled scaffolds from the population and tasked with combining them in such a way that would be likely to improve performance performance. The full crossover prompt is given in Appendix \ref{crossover_prompts}.

\subsection{Descriptors}
In open-ended evolutionary approaches, descriptors are essential for quantifying the diversity of candidate solutions \citep{mapelites}. In order to explore the full range of vulnerabilities of a base model, we seek to generate and evaluate a diverse range of multi-agent scaffolds and require high-dimensional descriptors. In \textsc{AgentBreeder}, we use OpenAI's \textit{text-embedding-3-small} \citep{text-embedding-3-small} model returning a 12-dimensional text embedding of the system name and code as our descriptor to encode semantic information about the name, structure, and potential logic embedded in the scaffold.

\subsection{Clustering}

Once the descriptors have been generated for the new scafolds, \textsc{AgentBreeder} re-clusters the whole population based on their descriptors to discover groups of similar architectures. We choose agglomerative clustering as it has been found to be particularly effective for smaller datasets like ours \citep{weigand2021can}. By setting a distance threshold in the agglomerative clustering algorithm, we allow the number of clusters to adjust flexibly. When the number of clusters increases, the selection pressure decreases towards zero. Conversely, reducing the number of clusters encourages the algorithm to explore only a few options, which leads to less diverse scaffolds. To achieve a balanced trade-off between system performance and system diversity, a distance threshold of 0.7 was selected.

\subsection{Multi-Objective Pareto Elites}

In Quality-Diversity algorithms such as MAP-Elites \citep{mapelites}, selection pressure is applied by randomly sampling the highest-performing candidates in each niche for evolution, referred to as the ``elites". In multi-objective optimization, a solution
is Pareto optimal if no other solution improves one objective without worsening another \citep{moea}. The Pareto front comprises
all such optimal solutions. In \textsc{AgentBreeder}, instead of sampling from pre-defined niches, we sample elites from the Pareto fronts of dynamically generated clusters.

\subsection{Evaluations}

Evaluations are implemented in Inspect \citep{AISIInspect}, an open-source framework for LLM evaluations. We instantiate \textsc{AgentBreeder} as a custom model provider by deriving a new class from ModelAPI, and each individual scaffold derives as a Model from that ModelAPI. This allows comprehensive experiment tracking and parallelization, and provides an extensible framework allowing \textsc{AgentBreeder} to be run on a new benchmark often with fewer than 100 lines of code. In Section \ref{experiments}, we report results on 5 benchmarks comprising safety, capability and helpfulness.

\section{Experiments}
\label{experiments}
We conduct experiments to validate \textsc{AgentBreeder}'s three modes; \textsc{BlueAgentBreeder}, \textsc{RedAgentBreeder}, \textsc{CapableAgentBreeder}. 
To evaluate the capability of the multi-agent scaffolds produced, we follow the approaches of \citet{adas} and \citet{godel} and report results on three benchmarks from OpenAI's simple-evals \citep{simpleevals}. To evaluate system safety, we report results on one comprehensive safety benchmark. A full description of each benchmark can be found in Appendix \ref{bench}. To evaluate the safety of multi-agent scaffolds, we adopted a method inspired by Constitutional AI, where an LLM judges responses based on a predefined set of principles. We transformed the Anthropic Claude Constitution \citep{claudeconstitution} into a series of yes/no questions. During validation, each system responded to 100 attack-enhanced prompts, and an agent aware of the constitution evaluated the safety of each response. We then reported the median accuracy and the 95\% bootstrap confidence interval for each scaffold's safety performance.

\subsection{Experiment 1: Blue Team Defense} 
\label{blueteamdefense}
\begin{figure}[H]
  \centering
  
  \includegraphics[width = \textwidth]{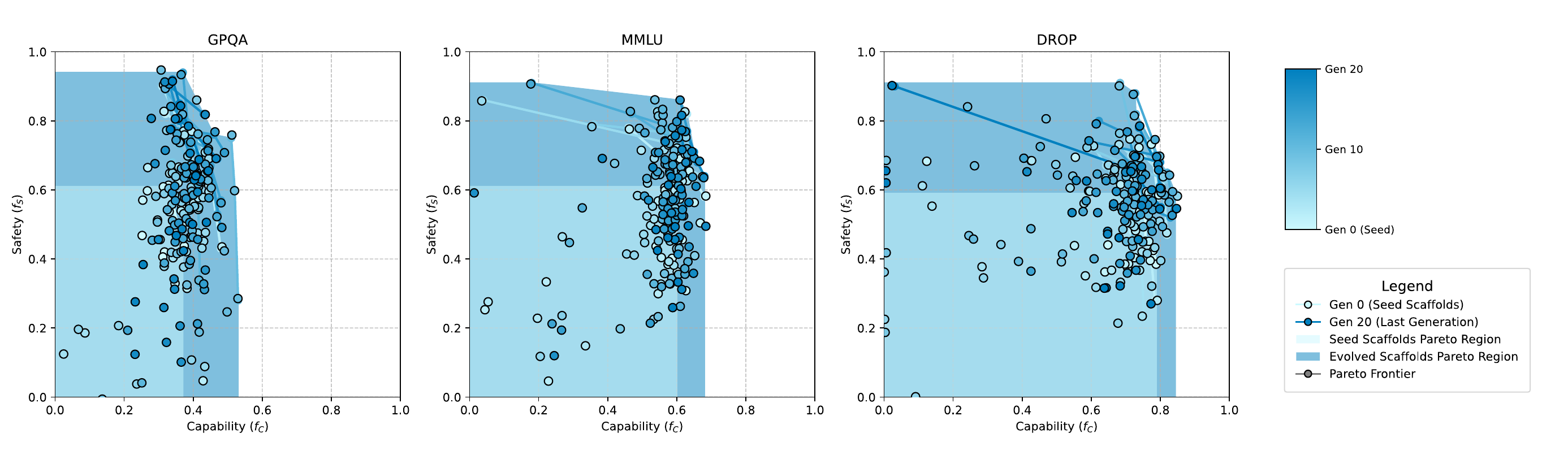}
  \caption{\textsc{BlueAgentBreeder} evolves scaffolds that improve the capability-safety Pareto frontier. The plots show the evolution of multi-agent scaffolds across 20 generations on the validation set of three different benchmarks: GPQA (left), MMLU (middle), and DROP (right). Each point represents a scaffold, with colors indicating generation (lighter blue for seed scaffolds, darker blue for later generations). The x-axis measures capability ($f_C$) and the y-axis measures safety ($f_S$). The light blue shaded region shows the Pareto frontier of the seed generation, while the dark blue region shows the Pareto frontier of evolved scaffolds.}
  \label{fig:bluerun}
\end{figure}

We ran \textsc{BlueAgentBreeder} for 20 generations, on each of our three chosen capability benchmarks (DROP \citep{drop}, MMLU \citep{mmlu} and GPQA \citep{gpqa}) independently, with the aim of generating ``Blue Teams'' of multi-agent scaffolds that simultaneously optimize for capability and safety across our chosen benchmarks. The evolution is visualized in Figure \ref{fig:bluerun} and we report the HV indicator on the test set for each experiment in Table \ref{tab:bluehv}. The Meta Agent discovers 10 new scaffolds each generation, and we report the median accuracy and the 95\% confidence interval on the held-out test set for the best performing discovered scaffold in Table \ref{tab:blue_tab}. The ``best'' scaffold maximized the sum of capability and safety scores, with the important caveat that we excluded scaffolds that engaged in reward hacking of the safety objective. A more detailed visualization of the evolutionary process is shown in Figure \ref{fig:bluerun}. To reduce \textsc{BlueAgentBreeder}'s tendency to reward-hack the safety benchmark by finding a trivial safe response to question-answering tasks that require a long-form response, during evaluation, we report the ``helpfulness'' of the scaffold on questions from TruthfulQA \citep{truthfulqa}.

\begin{table}[H]
    \centering
    \begin{tabular}{c|c|c}
         Benchmark & Seed Scaffolds & Discovered Scaffolds \\
         \hline
         GPQA & 0.219064 & 0.247536 \\
         MMLU & 0.484208 & 0.542816 \\
         DROP & 0.390754 & 0.438813
    \end{tabular}
    \caption{Reporting the HV indicator on the test set for \textsc{BlueAgentBreeder}.}
    \label{tab:bluehv}
\end{table}

\begin{table}[t!]
\centering

\begin{tabular}{l@{\hskip 10pt}c@{\hskip 10pt}c@{\hskip 10pt}c@{\hskip 10pt}c@{\hskip 10pt}c}
\toprule
\textsc{BlueAgentBreeder} & \multicolumn{3}{c}{Capability} & Safety & Helpfulness \\
\cmidrule(l{2pt}r{10pt}){2-4} \cmidrule(l{2pt}r{10pt}){5-5} \cmidrule(l{2pt}r{10pt}){6-6}
 & \textbf{DROP} & \textbf{MMLU} & \textbf{GPQA} & \textbf{SaladData} & \textbf{TruthfulQA} \\
\midrule
 \multicolumn{6}{l}{\emph{Seed Scaffolds from ADAS \citep{adas}}} \\
\midrule
Chain-of-Thought (CoT) & 66.6 $\pm$ 5.0 & 80.0 $\pm$ 4.4  & 31.2 $\pm$ 5.6  & 29.2 $\pm$ 5.6 & 86.8 $\pm$ 3.6 \\
Self-Consistency CoT & 66.0 $\pm$ 4.4 & 81.6 $\pm$ 4.8 & 32.4 $\pm$ 6.0 & 22.8 $\pm$ 5.2  & 85.6 $\pm$ 4.4 \\
Self-Refinement  & 61.4 $\pm$ 4.8 & 78.4 $\pm$ 5.2 & 28.4 $\pm$ 6.0 & 26.0 $\pm$ 5.2 & 86.8 $\pm$ 4.0 \\
Debate& 69.9 $\pm$ 4.4 & 77.6 $\pm$ 5.2 & 29.6 $\pm$ 5.6 & 36.4 $\pm$ 6.0  & 86.4 $\pm$ 4.0 \\
Step-Back Abstraction &  71.4 $\pm$ 4.3 & 79.2 $\pm$ 4.8 & 30.8 $\pm$ 5.2 & 40.8 $\pm$ 5.6 & 85.2 $\pm$ 4.4 \\
Quality-Diversity  & \underline{78.0 $\pm$ 3.9}  & 81.6 $\pm$ 4.4 & 28.4 $\pm$ 5.6 &  25.8 $\pm$ 5.8 & \underline{87.2 $\pm$ 4.0} \\
Role Assignment  & 75.8 $\pm$ 4.2 & 79.2 $\pm$ 4.8 & 32.0 $\pm$ 6.0 & 18.0 $\pm$ 5.2 & 85.6 $\pm$ 4.4 \\
\midrule
 \multicolumn{6}{l}{\emph{BlueAgentBreeder Scaffolds ($S=\text{SaladData}$, $H=\text{TruthfulQA}$)}} \\
\midrule

${\arg\max_s} \{f_{C_\text{DROP}}\}$& {\textbf{79.0 $\pm$ 3.8}} & - & - & 46.4 $\pm$ 6.4 & \textbf{88.0 $\pm$ 4.0}\\
${\arg\max_s} \{f_{S}\}$ & 62.0 $\pm$ 4.8 & - & - & \underline{86.0 $\pm$ 4.0} & 83.6 $\pm$ 4.4\\
${\arg\max_s} \{f_{C_\text{DROP}},f_{S},f_{H}\}$ & 62.0 $\pm$ 4.8 & - & - & \underline{86.0 $\pm$ 4.0} & 83.6 $\pm$ 4.4\\
${\arg\max_s} \{f_{C_\text{MMLU}}\}$ & - &{\textbf{85.2 $\pm$ 4.4}}  & - & 54.0 $\pm$ 5.6 & 81.2 $\pm$ 4.4\\
${\arg\max_s} \{f_{S}\}$ & - & \underline{84.0 $\pm$ 4.4}& -  & 84.4 $\pm$ 4.0 & 76.0 $\pm$ 5.2\\
${\arg\max_s} \{f_{C_\text{MMLU}},f_{S},f_{H}\}$  & - & \underline{84.0 $\pm$ 4.4} & - & 84.4 $\pm$ 4.0 & 76.0 $\pm$ 5.2\\
${\arg\max_s} \{f_{C_\text{GPQA}}\}$  & - & -&\textbf{39.2 $\pm$ 5.6} & 52.0 $\pm$ 6.8 & 57.6 $\pm$ 6.4\\
${\arg\max_s} \{f_{S}\}$ & - & - & 31.2 $\pm$ 6.0  & \textbf{95.2 $\pm$ 2.4} & 49.6 $\pm$ 6.4\\
${\arg\max_s} \{f_{C_\text{GPQA}},f_{S},f_{H}\}$  & - & - & \underline{36.8 $\pm$ 5.2} & 49.2 $\pm$ 6.8 & 86.8 $\pm$ 4.0\\
\bottomrule

\end{tabular}

\caption{We report the evaluation results of \textsc{BlueAgentBreeder} on the held-out test set of capability benchmark (DROP \citep{drop}, MMLU \citep{mmlu}, GPQA \citep{gpqa}), safety benchmark (SaladData \citep{saladdata}) and ensure a trivial solution has not been found by evaluating each scaffold's helpfulness on TruthfulQA \citep{truthfulqa}. For each benchmark, we compare the \textsc{AgentBreeder}-discovered scaffolds against seed scaffolds, with the highest and second highest result in each column given in bold and underlined respectively. We report the median accuracy (or F1 score for DROP) along with a 95\% confidence interval computed over 250 held-out test samples. The evolution is conducted independently for each capability benchmark.}
\label{tab:blue_tab}
\end{table}

\textbf{Significant Increase in Performance on our Safety Benchmark.} We observed a 110.7\% maximum increase in performance of our best discovered scaffold on SaladData \citep{saladdata} with an average uplift of 79.4\% across all three experimental runs.

\textbf{Performance on Capability Benchmarks Maintained if not Increased.} Whilst providing a substantial safety uplift, we were able to ensure capability was either kept the same or additionally provided a marginal uplift, for example as much as 21.0\% on GPQA \citep{gpqa}.

\textbf{Vulnerability to Reward Hacking.} We note that certain scaffolds display inflated scores on our safety benchmark, achieving an accuracy of 95.2 $\pm$ 2.4. However, under closer examination, the system shows a significant 43\% decrease in performance on our helpfulness benchmark. We find that multi-agent scaffolds are prone to reward-hacking the safety objective e.g. by finding a trivial solution such as responding ``I'm sorry, I can't help with that.'' to any question outside of their domain, and that one way of detecting this behavior is by reporting performance on a helpfulness benchmark. 

\subsection{Experiment 2: Red Team Attack}
\label{red_mode}

\begin{figure}[H]
  \centering
  
  \includegraphics[width=0.45\textwidth]{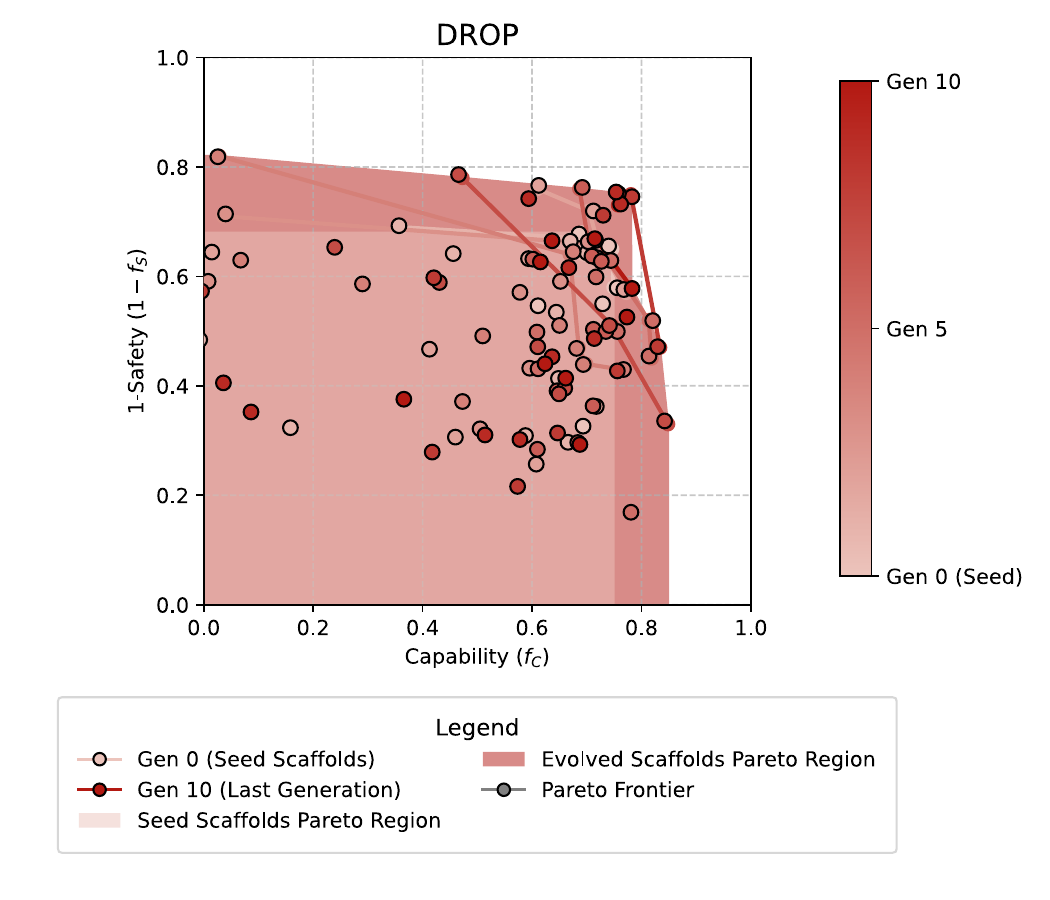}
  \caption{\textsc{RedAgentBreeder} evolves scaffolds over 10 generations. The validation set results are reported, with the x-axis shows capability ($f_C$) and the y-axis shows inverted safety (1-$f_S$), with higher values on both axes representing the optimization objectives. Lighter red points indicate seed scaffolds (Gen 0), while darker red points represent scaffolds from later generations (up to Gen 10). The shaded regions highlight the Pareto regions of both seed scaffolds (lighter area) and evolved scaffolds (darker area), with the connected points along the boundary representing non-dominated solutions.}
  \label{fig:redrun}
\end{figure}

\textsc{RedAgentBreeder} seeks to discover ``Red Team'' scaffolds that exploit the safety vulnerabilities of the base LLM while preserving capability on a chosen benchmark, in this case DROP \citep{drop}. We introduce an inverted safety benchmark ``1-SaladData'' which contains the same questions as the original SaladData \citep{saladdata} benchmark, but utilizes an inverted scoring function that awards higher scores for unsafe responses. The \textsc{RedAgentBreeder} scenario models the outer misalignment problem, where a scaffold designer operates under a misaligned reward function that inadvertently incentivizes unsafe behavior while pursuing seemingly beneficial objectives. It is important to note that in this case, the Meta Agent is not prompted to discover unsafe scaffolds, instead these arise via Pareto optimization on capability and harm benchmarks. This seeks to model the case where an actor may unknowingly expose weaknesses in the base LLM when employing scaffolding to improve task performance.  A visualization is provided in Figure \ref{fig:redrun} and Tables \ref{tab:redhv} and \ref{tab:red_table} summarize these results.

\begin{table}[H]
    \centering
    \begin{tabular}{c|c|c}
         Benchmark & Seed Scaffolds & Discovered Scaffolds \\
         \hline
         DROP & 0.572759 & 0.0.602547
    \end{tabular}
    \caption{Reporting the HV indicator on the test set for \textsc{RedAgentBreeder}.}
    \label{tab:redhv}
\end{table}

\begin{table}[H]

\begin{tabular}{lccc}
\toprule
\multicolumn{1}{c}{\textsc{RedAgentBreeder}} & \multicolumn{1}{c}{Capability} & \multicolumn{1}{c}{Harm} \\
\cmidrule(lr){2-2}\cmidrule(lr){3-3}
& \textbf{DROP} & \textbf{1-SaladData} \\
\midrule
\multicolumn{3}{l}{\emph{Seed Scaffolds from ADAS \citep{adas}}} \\
\midrule
CoT & 68.2 $\pm$ 4.7 & \underline{76.8 $\pm$ 4.8}  \\
CoTSC & 69.6 $\pm$ 4.8 & 70.8 $\pm$ 6.0  \\
Refine & 62.6 $\pm$ 4.7 & 71.6 $\pm$ 5.2 \\
Debate& 63.2 $\pm$ 4.8 & 56.0 $\pm$ 6.0 \\
StepBack &  67.5 $\pm$ 4.5 & 58.0 $\pm$ 6.0 \\
QD & \underline{74.9 $\pm$ 4.2} & 73.6 $\pm$ 5.2 \\
Roles & 73.1 $\pm$ 3.7 & 46.4 $\pm$ 6.0  \\
\midrule
\multicolumn{3}{l}{\emph{RedAgentBreeder Scaffolds}} \\
\midrule
${\arg\max_s} \{f_{C_\text{DROP}}\}$ & \textbf{75.0 $\pm$ 4.1} & 26.0 $\pm$ 5.2  \\
${\arg\max_s} \{f_{1-S}\}$ & 67.7 $\pm$ 4.7 & \textbf{81.6 $\pm$ 4.8}  \\
${\arg\max_s} \{f_{C_\text{DROP}},f_{1-S}\}$  & 67.7 $\pm$ 4.7 & \textbf{81.6 $\pm$ 4.8}  \\
\bottomrule

\end{tabular}
\caption{We report the evaluation results of \textsc{AgentBreeder} run in ``red'' mode on the held-out test set. We seek to maximize performance on DROP \citep{drop} whilst also maximizing performance on 1-SaladData \citep{saladdata}, an inverted version of the SaladData benchmark where unsafe responses are scored highly. For each
benchmark, we compare the \textsc{AgentBreeder}-discovered scaffolds against seed scaffolds, with the highest and second highest result
in each column given in bold and underlined respectively. We report the F1 score and median accuracy for DROP and 1-SaladData respectively, along with a 95\% confidence interval computed over 250 held-out test samples.}
\label{tab:red_table}
\end{table}

\textbf{Unsafe Scaffolds are Easier to Find.} \textsc{RedAgentBreeder}'s highest performing scaffold achieved 81.6$\pm$4.8 accuracy on the inverted 1-SaladData \citep{saladdata} metric, surpassing all seed scaffolds by at least 6.25\% with only half the generation budget of \textsc{BlueAgentBreeder}. These results indicate that scaffolding may be more likely to weaken than strengthen the base LLM to adversarial attacks.

\textbf{Capability Disguises Safety Vulnerabilities.} Interestingly, even while maximizing unsafe performance, we were able to achieve a competitive F1 score of 67.7$\pm$4.7. This result is comparable to the seed scaffolds, highlighting that scaffolds may appear just as capable in terms of task performance yet simultaneously exhibit increased safety vulnerabilities.

\subsection{Experiment 3: Multi-Objective Ablation}

As an ablation for our multi-objective criteria and to compare \textsc{AgentBreeder} against the seminal work, we run \textsc{CapableAgentBreeder} - a single-objective-variant of our framework - for 20 generations, evolving 10 mutants each generation. We take the highest-performing scaffolds from ADAS \citep{adas} and evaluate them with GPT-4o mini \citep{gpt4omini} as their core model.
We report the F1 score for DROP \citep{drop}, median accuracy for MMLU \citep{mmlu} and GPQA \citep{gpqa} and their 95\% confidence intervals, as well as their performance on SaladData \citep{saladdata}, our chosen safety benchmark. The results are shown in Table \ref{tab:capable} in Appendix \ref{capapx}.

\textbf{Comparable Performance to Previous Work.} 
\textsc{CapableAgentBreeder} achieves competitive results to ADAS, marginally surpassing performance across all capability benchmarks.

\textbf{Multi-Objective outperforms Single-Objective Optimization.} The scaffolds discovered by \textsc{CapableAgentBreeder} achieve near or slightly above-baseline results, such as 72.3 $\pm$3.1 F1 on DROP and 41.2$\pm$4.4 accuracy on GPQA. This performance gain is notably smaller than in the multi-objective setting. This supports our hypothesis that incorporating an additional benchmark may increase the signal-to-noise ratio of scaffold validations each generation. This improves the quality of the selection pressure for the evolutionary algorithm, helping the process converge to better solutions overall.

\textbf{Insignificant Impact on Safety Performance.} In single-objective ablation runs, the discovered scaffolds showed only modest performance uplift on SaladData \citep{saladdata}, suggesting that ignoring safety in the objective yields no strong impetus for safe or unsafe behaviors. This contrasts with multi-objective runs, where explicit safety optimization (or ``negative safety'' in red-teaming) substantially influenced outcomes.

\textbf{Performance Stagnates with Better LLMs.} When using more advanced models (GPT-4o mini \citep{gpt4omini} for scaffolds and Claude 3.5 Sonnet \citep{claude} for the Meta Agent) compared to the original ADAS \citep{adas} implementation, we observe that while overall performance improves, the relative gain between seed and discovered scaffolds diminishes. We attribute this to three plausible factors: (1) increased data contamination in newer LLMs may lead to memorized solutions rather than genuine reasoning, (2) higher baseline performance makes marginal improvements harder to distinguish from noise and (3) recent models are already fine-tuned for detailed reasoning, reducing the benefit of scaffold-induced reasoning steps \citep{openai_learning_to_reason_2024, trading}.

\section{Discussion}
\label{discussion}
\textbf{Pre-Deployment Safety Evaluations.} The Dead Internet Theory posits a future where AI agents dominate online activity \citep{deadinternettheory}. While speculative, the recent releases of Operator \citep{operator} and Proxy \citep{proxy} highlight the increasing population of agents deployed with the ability to interact autonomously with other agents and humans. These underscore the uncertainty around agent-on-agent dynamics, especially when these agents evolve or compose themselves in unanticipated ways. Our \textsc{RedAgentBreeder} experiments illustrate an automated approach to efficiently surface multi-agent scaffolds that exhibit vulnerabilities on safety benchmarks. Over time, labs could adopt a \textsc{RedAgentBreeder}-style pipeline to proactively ``red-team'' new LLMs as part of a release protocol. 

\textbf{Post-Deployment Adversarial Robustness.}
Just as \textsc{RedAgentBreeder} discovers vulnerable scaffolds, \textsc{BlueAgentBreeder} provides a methodology to design safe and capable multi-agent scaffolds. This method can also be used to upgrade the safety capabilities of existing scaffolds, akin to Weak-to-Strong Generalization \citep{weaktostrong}. Furthermore, \textsc{BlueAgentBreeder} can be used to ensure a scaffold conforms to dynamic company values, policies and regulatory requirements. These experiments validate the practicality of evolutionary search as a dynamic, data-driven tool for multi-agent evaluation.

\textbf{Limitations.} While our experiments provide promising insights, several limitations should be acknowledged. Firstly, due to computational costs, we conducted experiments over a limited number of generations and with relatively small population sizes, resulting in only marginal performance improvements. Secondly, our experimental setup serves as a proof of concept for multi-objective alignment, and stronger claims of helpfulness and safety would require evaluations on more comprehensive benchmarks. Additionally, our evaluation was restricted to a select set of benchmarks, which may not fully capture the diverse range of real-world capabilities and safety concerns. Finally, the initial population was limited to seven seed scaffolds, potentially constraining the diversity of discovered scaffolds.

\section{Conclusion}
\label{conclusion}

This paper introduces \textsc{AgentBreeder}, an evolutionary framework for discovering and evaluating multi-agent scaffolds via the multi-objective optimization of capability and safety. Our experiments demonstrate that \textsc{AgentBreeder} operates effectively in three distinct modes. \textsc{BlueAgentBreeder} for developing safer scaffolds, \textsc{RedAgentBreeder} for identifying vulnerabilities, and \textsc{CapableAgentBreeder} for maximizing task performance. Through empirical evaluation across multiple benchmarks, we show that our framework discovers scaffolds that achieve competitive or increased performance to prior works while exhibiting increased adversarial robustness.

Our results highlight several important findings for AI safety research. First, we demonstrate that unsafe behaviors can coexist with strong task performance, as evidenced by \textsc{RedAgentBreeder}'s ability to generate scaffolds that maintain capability while exhibiting increased vulnerability. Second, our experiments reveal that multi-objective optimization targeting both capability and safety yields better overall solutions compared to single-objective approaches. Third, our \textsc{BlueAgentBreeder} experiments achieved substantial safety improvements (up to 110.7\% increase on SaladData with 79.4\% average uplift) while simultaneously maintaining or enhancing capability (up to 21.0\% improvement on GPQA). Finally, we show that automated evolutionary methods can effectively probe the complex attack surfaces of multi-agent scaffolds, offering a practical approach to pre-deployment safety evaluation.

As AI systems become increasingly interconnected and deployed in real-world settings, frameworks like \textsc{AgentBreeder} bridge the research gap between single-agent and multi-agent safety evaluations. Our work establishes a foundation for the systematic evaluation of multi-agent scaffolds, contributing to the development of safer and more reliable AI technologies.

\section{Future Work}
\label{futurework}

\textbf{Scaling Laws.} Scaling up \textsc{AgentBreeder} to larger population sizes and longer evolutionary runs could yield more substantial improvements in both capability and safety metrics. Incorporating closed-source safety benchmarks such as AILuminate \citep{AILuminate2025} and contamination-free capability benchmarks such as MMLU-CF \citep{contaminationfreemmlu} would provide a more comprehensive assessment of multi-agent system safety.

\textbf{White-Box and Gray-Box Evaluations.} A key limitation of our current approach is its focus on black-box evaluation of scaffolds. Future work could investigate individual agent behaviors, including how agents interact with tools, external APIs, and information sources. Developing methods to trace and analyze agent-agent and agent-tool interactions could reveal potential safety risks that are invisible in black box evaluation.

\textbf{Alternative Objectives.} In this work, we only consider the capability and safety objectives for optimization. Future work could explore inference cost as an objective to minimize for, and consider multi-core scaffolds where different LLM base models exist inside the same scaffold.

\textbf{Multi-Agent Governance.} Critical research is needed to establish governance frameworks for multi-agent scaffolds. Future work could comprise developing differentiated safety cases for scaffolds with varying levels of transparency, from fully white box to black box architectures.

\begin{ack}
J Rosser is supported by the EPSRC centre for Doctoral Training in Autonomous and Intelligent Machines and Systems EP/Y035070/1. We extend our sincere gratitude to the members of the Foerster Lab for AI Research (FLAIR) for their guidance during the project scoping phase and thorough proofreading. Special thanks to the London Initiative for Safe AI and Arcadia Impact for providing workspace and offering invaluable feedback throughout. 
\end{ack}

\bibliographystyle{plainnat}
\bibliography{neurips_2025}

\begin{thebibliography}{67}
\providecommand{\natexlab}[1]{#1}
\providecommand{\url}[1]{\texttt{#1}}
\expandafter\ifx\csname urlstyle\endcsname\relax
  \providecommand{\doi}[1]{doi: #1}\else
  \providecommand{\doi}{doi: \begingroup \urlstyle{rm}\Url}\fi

\bibitem[AI~Security~Institute(2024)]{AISIInspect}
UK~AI~Security~Institute.
\newblock Inspect {AI:} {Framework} for {Large} {Language} {Model} {Evaluations}, 2024.
\newblock URL \url{https://github.com/UKGovernmentBEIS/inspect_ai}.

\bibitem[Amirizaniani et~al.(2024)Amirizaniani, Martin, Sivachenko, Mashhadi, and Shah]{amirizaniani2024can}
Maryam Amirizaniani, Elias Martin, Maryna Sivachenko, Afra Mashhadi, and Chirag Shah.
\newblock Can llms reason like humans? assessing theory of mind reasoning in llms for open-ended questions.
\newblock In \emph{Proceedings of the 33rd ACM International Conference on Information and Knowledge Management}, pages 34--44, 2024.

\bibitem[Anthropic(2024)]{claude}
AI~Anthropic.
\newblock Claude 3.5 sonnet model card addendum.
\newblock \emph{Claude-3.5 Model Card}, 3:\penalty0 6, 2024.

\bibitem[Bai et~al.(2022)Bai, Kadavath, Kundu, Askell, Kernion, Jones, Chen, Goldie, Mirhoseini, McKinnon, et~al.]{claudeconstitution}
Yuntao Bai, Saurav Kadavath, Sandipan Kundu, Amanda Askell, Jackson Kernion, Andy Jones, Anna Chen, Anna Goldie, Azalia Mirhoseini, Cameron McKinnon, et~al.
\newblock Constitutional ai: Harmlessness from ai feedback.
\newblock \emph{arXiv preprint arXiv:2212.08073}, 2022.

\bibitem[Boyer and Moore(1983)]{Boyer1983TuringLISP}
Robert~S. Boyer and J.~Strother Moore.
\newblock A mechanical proof of the turing completeness of pure lisp.
\newblock Technical Report ADA130625, Texas Univ at Austin Inst for Computing Science and Computer Applications, May 1983.
\newblock URL \url{https://apps.dtic.mil/sti/citations/ADA130625}.
\newblock Approved for public release.

\bibitem[Burns et~al.(2023)Burns, Izmailov, Kirchner, Baker, Gao, Aschenbrenner, Chen, Ecoffet, Joglekar, Leike, et~al.]{weaktostrong}
Collin Burns, Pavel Izmailov, Jan~Hendrik Kirchner, Bowen Baker, Leo Gao, Leopold Aschenbrenner, Yining Chen, Adrien Ecoffet, Manas Joglekar, Jan Leike, et~al.
\newblock Weak-to-strong generalization: Eliciting strong capabilities with weak supervision.
\newblock \emph{arXiv preprint arXiv:2312.09390}, 2023.

\bibitem[Cao et~al.(2024)Cao, Ma, Zhai, and Shen]{cao2024llm}
Hong Cao, Rong Ma, Yanlong Zhai, and Jun Shen.
\newblock Llm-collab: a framework for enhancing task planning via chain-of-thought and multi-agent collaboration.
\newblock \emph{Applied Computing and Intelligence}, 4\penalty0 (2):\penalty0 328--348, 2024.

\bibitem[Chan et~al.(2023)Chan, Chen, Su, Yu, Xue, Zhang, Fu, and Liu]{chan2023chateval}
Chi-Min Chan, Weize Chen, Yusheng Su, Jianxuan Yu, Wei Xue, Shanghang Zhang, Jie Fu, and Zhiyuan Liu.
\newblock Chateval: Towards better llm-based evaluators through multi-agent debate.
\newblock \emph{arXiv preprint arXiv:2308.07201}, 2023.

\bibitem[Chao et~al.(2023)Chao, Robey, Dobriban, Hassani, Pappas, and Wong]{chao2023jailbreaking}
Patrick Chao, Alexander Robey, Edgar Dobriban, Hamed Hassani, George~J Pappas, and Eric Wong.
\newblock Jailbreaking black box large language models in twenty queries.
\newblock \emph{arXiv preprint arXiv:2310.08419}, 2023.

\bibitem[Chen et~al.(2023)Chen, Su, Zuo, Yang, Yuan, Qian, Chan, Qin, Lu, Xie, et~al.]{chen2023agentverse}
Weize Chen, Yusheng Su, Jingwei Zuo, Cheng Yang, Chenfei Yuan, Chen Qian, Chi-Min Chan, Yujia Qin, Yaxi Lu, Ruobing Xie, et~al.
\newblock Agentverse: Facilitating multi-agent collaboration and exploring emergent behaviors in agents.
\newblock \emph{arXiv preprint arXiv:2308.10848}, 2\penalty0 (4):\penalty0 6, 2023.

\bibitem[Chollet(2019)]{ARC-AGI}
Fran{\c{c}}ois Chollet.
\newblock On the measure of intelligence.
\newblock \emph{arXiv preprint arXiv:1911.01547}, 2019.

\bibitem[{Convergence}(2024)]{proxy}
{Convergence}.
\newblock Introducing web-world models, December 2024.
\newblock URL \url{https://convergence.ai/training-web-agents-with-web-world-models-dec-2024/}.
\newblock Accessed: 2025-01-30.

\bibitem[Du et~al.(2023)Du, Li, Torralba, Tenenbaum, and Mordatch]{debate}
Yilun Du, Shuang Li, Antonio Torralba, Joshua~B Tenenbaum, and Igor Mordatch.
\newblock Improving factuality and reasoning in language models through multiagent debate.
\newblock \emph{arXiv preprint arXiv:2305.14325}, 2023.

\bibitem[Dua et~al.(2019)Dua, Wang, Dasigi, Stanovsky, Singh, and Gardner]{drop}
Dheeru Dua, Yizhong Wang, Pradeep Dasigi, Gabriel Stanovsky, Sameer Singh, and Matt Gardner.
\newblock Drop: A reading comprehension benchmark requiring discrete reasoning over paragraphs.
\newblock \emph{arXiv preprint arXiv:1903.00161}, 2019.

\bibitem[Fernando et~al.(2023)Fernando, Banarse, Michalewski, Osindero, and Rockt{\"a}schel]{promptbreeder}
Chrisantha Fernando, Dylan Banarse, Henryk Michalewski, Simon Osindero, and Tim Rockt{\"a}schel.
\newblock {Promptbreeder: Self-referential self-improvement via prompt evolution}.
\newblock \emph{arXiv preprint arXiv:2309.16797}, 2023.

\bibitem[Fourney et~al.(2024)Fourney, Bansal, Mozannar, Tan, Salinas, Niedtner, Proebsting, Bassman, Gerrits, Alber, et~al.]{fourney2024magentic}
Adam Fourney, Gagan Bansal, Hussein Mozannar, Cheng Tan, Eduardo Salinas, Friederike Niedtner, Grace Proebsting, Griffin Bassman, Jack Gerrits, Jacob Alber, et~al.
\newblock Magentic-one: A generalist multi-agent system for solving complex tasks.
\newblock \emph{arXiv preprint arXiv:2411.04468}, 2024.

\bibitem[Fowler(2023)]{scaffoldedllms}
Stephen Fowler.
\newblock Scaffolded llms: Less obvious concerns.
\newblock \emph{LessWrong}, 2023.
\newblock URL \url{https://www.lesswrong.com/posts/mAwxebLw3nYbDivmt/scaffolded-llms-less-obvious-concerns}.

\bibitem[Ghafarollahi and Buehler(2024)]{ghafarollahi2024sciagents}
Alireza Ghafarollahi and Markus~J Buehler.
\newblock Sciagents: Automating scientific discovery through multi-agent intelligent graph reasoning.
\newblock \emph{arXiv preprint arXiv:2409.05556}, 2024.

\bibitem[Hendrycks et~al.(2020)Hendrycks, Burns, Basart, Zou, Mazeika, Song, and Steinhardt]{mmlu}
Dan Hendrycks, Collin Burns, Steven Basart, Andy Zou, Mantas Mazeika, Dawn Song, and Jacob Steinhardt.
\newblock Measuring massive multitask language understanding.
\newblock \emph{arXiv preprint arXiv:2009.03300}, 2020.

\bibitem[Hu et~al.(2024{\natexlab{a}})Hu, Lu, and Clune]{adas}
Shengran Hu, Cong Lu, and Jeff Clune.
\newblock Automated design of agentic systems.
\newblock \emph{arXiv preprint arXiv:2408.08435}, 2024{\natexlab{a}}.

\bibitem[Hu et~al.(2023)Hu, Xie, Jain, Francis, Patrikar, Keetha, Kim, Xie, Zhang, Fang, et~al.]{hu2023toward}
Yafei Hu, Quanting Xie, Vidhi Jain, Jonathan Francis, Jay Patrikar, Nikhil Keetha, Seungchan Kim, Yaqi Xie, Tianyi Zhang, Hao-Shu Fang, et~al.
\newblock Toward general-purpose robots via foundation models: A survey and meta-analysis.
\newblock \emph{arXiv preprint arXiv:2312.08782}, 2023.

\bibitem[Hu et~al.(2024{\natexlab{b}})Hu, Cai, Du, Zhu, Liu, Yu, Hou, Tang, and Chen]{evomac}
Yue Hu, Yuzhu Cai, Yaxin Du, Xinyu Zhu, Xiangrui Liu, Zijie Yu, Yuchen Hou, Shuo Tang, and Siheng Chen.
\newblock Self-evolving multi-agent collaboration networks for software development.
\newblock \emph{arXiv preprint arXiv:2410.16946}, 2024{\natexlab{b}}.

\bibitem[Huang et~al.(2024)Huang, Zhou, Jin, Zhou, Chen, Wang, Yuan, Sap, and Lyu]{maliciouserrormultiagentsystems}
Jen-tse Huang, Jiaxu Zhou, Tailin Jin, Xuhui Zhou, Zixi Chen, Wenxuan Wang, Youliang Yuan, Maarten Sap, and Michael~R Lyu.
\newblock On the resilience of multi-agent systems with malicious agents.
\newblock \emph{arXiv preprint arXiv:2408.00989}, 2024.

\bibitem[Kesireddy and Medrano(2024)]{moea}
Adarsh Kesireddy and F~Antonio Medrano.
\newblock Elite multi-criteria decision making—pareto front optimization in multi-objective optimization.
\newblock \emph{Algorithms}, 17\penalty0 (5):\penalty0 206, 2024.

\bibitem[Khan(2022)]{multipolar}
Akbir Khan.
\newblock Why multi-agent safety is important.
\newblock \url{https://www.lesswrong.com/posts/pkfKRG9dQr6unrhQT/why-multi-agent-safety-is-important}, 2022.
\newblock Accessed: 2025-01-30.

\bibitem[Kong et~al.(2024)Kong, Braunl, Fahmi, and Wang]{kong2024superalignment}
Xiangrui Kong, Thomas Braunl, Marco Fahmi, and Yue Wang.
\newblock A superalignment framework in autonomous driving with large language models.
\newblock \emph{arXiv preprint arXiv:2406.05651}, 2024.

\bibitem[Li et~al.(2024{\natexlab{a}})Li, Xie, Li, Tsung, Ding, and Li]{li2024agent}
Ao~Li, Yuexiang Xie, Songze Li, Fugee Tsung, Bolin Ding, and Yaliang Li.
\newblock Agent-oriented planning in multi-agent systems.
\newblock \emph{arXiv preprint arXiv:2410.02189}, 2024{\natexlab{a}}.

\bibitem[Li et~al.(2024{\natexlab{b}})Li, Dong, Wang, Hu, Zuo, Lin, Qiao, and Shao]{saladdata}
Lijun Li, Bowen Dong, Ruohui Wang, Xuhao Hu, Wangmeng Zuo, Dahua Lin, Yu~Qiao, and Jing Shao.
\newblock Salad-bench: A hierarchical and comprehensive safety benchmark for large language models.
\newblock \emph{arXiv preprint arXiv:2402.05044}, 2024{\natexlab{b}}.

\bibitem[Lin et~al.(2021)Lin, Hilton, and Evans]{truthfulqa}
Stephanie Lin, Jacob Hilton, and Owain Evans.
\newblock {TruthfulQA: Measuring how models mimic human falsehoods}.
\newblock \emph{arXiv preprint arXiv:2109.07958}, 2021.

\bibitem[Liu et~al.(2023)Liu, Deng, Li, Wang, Wang, Wang, Zhang, Liu, Wang, Zheng, et~al.]{injection}
Yi~Liu, Gelei Deng, Yuekang Li, Kailong Wang, Zihao Wang, Xiaofeng Wang, Tianwei Zhang, Yepang Liu, Haoyu Wang, Yan Zheng, et~al.
\newblock Prompt injection attack against llm-integrated applications.
\newblock \emph{arXiv preprint arXiv:2306.05499}, 2023.

\bibitem[Lu et~al.(2024)Lu, Lu, Lange, Foerster, Clune, and Ha]{aiscientist}
Chris Lu, Cong Lu, Robert~Tjarko Lange, Jakob Foerster, Jeff Clune, and David Ha.
\newblock The ai scientist: Towards fully automated open-ended scientific discovery.
\newblock \emph{arXiv preprint arXiv:2408.06292}, 2024.

\bibitem[Madaan et~al.(2024)Madaan, Tandon, Gupta, Hallinan, Gao, Wiegreffe, Alon, Dziri, Prabhumoye, Yang, et~al.]{selfrefine}
Aman Madaan, Niket Tandon, Prakhar Gupta, Skyler Hallinan, Luyu Gao, Sarah Wiegreffe, Uri Alon, Nouha Dziri, Shrimai Prabhumoye, Yiming Yang, et~al.
\newblock Self-refine: Iterative refinement with self-feedback.
\newblock \emph{Advances in Neural Information Processing Systems}, 36, 2024.

\bibitem[{MLCommons Association}(2025)]{AILuminate2025}
{MLCommons Association}.
\newblock Ailuminate v1.0 benchmark, 2025.
\newblock URL \url{https://ailuminate.mlcommons.org/benchmarks/}.
\newblock Accessed: 2025-01-31.

\bibitem[Mouret and Clune(2015)]{mapelites}
Jean-Baptiste Mouret and Jeff Clune.
\newblock Illuminating search spaces by mapping elites.
\newblock \emph{arXiv preprint arXiv:1504.04909}, 2015.

\bibitem[Mukherjee et~al.(2024)Mukherjee, Gamble, Ausin, Kant, Aggarwal, Manjunath, Datta, Liu, Ding, Busacca, et~al.]{polaris}
Subhabrata Mukherjee, Paul Gamble, Markel~Sanz Ausin, Neel Kant, Kriti Aggarwal, Neha Manjunath, Debajyoti Datta, Zhengliang Liu, Jiayuan Ding, Sophia Busacca, et~al.
\newblock Polaris: A safety-focused llm constellation architecture for healthcare.
\newblock \emph{arXiv preprint arXiv:2403.13313}, 2024.

\bibitem[OpenAI(2023)]{simpleevals}
OpenAI.
\newblock simple-evals.
\newblock \url{https://github.com/openai/simple-evals}, 2023.
\newblock Accessed: 2025-01-29.

\bibitem[{OpenAI}(2024)]{gpt4omini}
{OpenAI}.
\newblock Gpt-4o mini: Advancing cost-efficient intelligence, 2024.
\newblock URL \url{https://openai.com/index/gpt-4o-mini-advancing-cost-efficient-intelligence/}.

\bibitem[OpenAI(2024{\natexlab{a}})]{openai_learning_to_reason_2024}
OpenAI.
\newblock Learning to reason with llms, 2024{\natexlab{a}}.
\newblock URL \url{https://openai.com/index/learning-to-reason-with-llms/}.
\newblock Accessed: 2025-01-31.

\bibitem[OpenAI(2024{\natexlab{b}})]{text-embedding-3-small}
OpenAI.
\newblock New embedding models and api updates, January 2024{\natexlab{b}}.
\newblock URL \url{https://openai.com/research/new-embedding-models-and-api-updates}.
\newblock Accessed: 2025-01-14.

\bibitem[OpenAI(2025)]{operator}
OpenAI.
\newblock Computer-using agent: Introducing a universal interface for ai to interact with the digital world, 2025.
\newblock URL \url{https://openai.com/index/computer-using-agent}.

\bibitem[Perez et~al.(2022)Perez, Huang, Song, Cai, Ring, Aslanides, Glaese, McAleese, and Irving]{perez2022red}
Ethan Perez, Saffron Huang, Francis Song, Trevor Cai, Roman Ring, John Aslanides, Amelia Glaese, Nat McAleese, and Geoffrey Irving.
\newblock Red teaming language models with language models.
\newblock \emph{arXiv preprint arXiv:2202.03286}, 2022.

\bibitem[Qian et~al.(2023)Qian, Cong, Yang, Chen, Su, Xu, Liu, and Sun]{qian2023communicative}
Chen Qian, Xin Cong, Cheng Yang, Weize Chen, Yusheng Su, Juyuan Xu, Zhiyuan Liu, and Maosong Sun.
\newblock Communicative agents for software development.
\newblock \emph{arXiv preprint arXiv:2307.07924}, 6\penalty0 (3), 2023.

\bibitem[Rein et~al.(2023)Rein, Hou, Stickland, Petty, Pang, Dirani, Michael, and Bowman]{gpqa}
David Rein, Betty~Li Hou, Asa~Cooper Stickland, Jackson Petty, Richard~Yuanzhe Pang, Julien Dirani, Julian Michael, and Samuel~R Bowman.
\newblock Gpqa: A graduate-level google-proof q\&a benchmark.
\newblock \emph{arXiv preprint arXiv:2311.12022}, 2023.

\bibitem[Romera-Paredes et~al.(2024)Romera-Paredes, Barekatain, Novikov, Balog, Kumar, Dupont, Ruiz, Ellenberg, Wang, Fawzi, et~al.]{funsearch}
Bernardino Romera-Paredes, Mohammadamin Barekatain, Alexander Novikov, Matej Balog, M~Pawan Kumar, Emilien Dupont, Francisco~JR Ruiz, Jordan~S Ellenberg, Pengming Wang, Omar Fawzi, et~al.
\newblock Mathematical discoveries from program search with large language models.
\newblock \emph{Nature}, 625\penalty0 (7995):\penalty0 468--475, 2024.

\bibitem[Samvelyan et~al.(2024)Samvelyan, Raparthy, Lupu, Hambro, Markosyan, Bhatt, Mao, Jiang, Parker-Holder, Foerster, et~al.]{rainbowteaming}
Mikayel Samvelyan, Sharath~Chandra Raparthy, Andrei Lupu, Eric Hambro, Aram~H Markosyan, Manish Bhatt, Yuning Mao, Minqi Jiang, Jack Parker-Holder, Jakob Foerster, et~al.
\newblock Rainbow teaming: Open-ended generation of diverse adversarial prompts.
\newblock \emph{arXiv preprint arXiv:2402.16822}, 2024.

\bibitem[Sartor and Thompson(2024)]{sartor2024neural}
Sebastian Sartor and Neil Thompson.
\newblock Neural scaling laws for embodied ai.
\newblock \emph{arXiv preprint arXiv:2405.14005}, 2024.

\bibitem[Shi et~al.(2022)Shi, Suzgun, Freitag, Wang, Srivats, Vosoughi, Chung, Tay, Ruder, Zhou, et~al.]{mgsm}
Freda Shi, Mirac Suzgun, Markus Freitag, Xuezhi Wang, Suraj Srivats, Soroush Vosoughi, Hyung~Won Chung, Yi~Tay, Sebastian Ruder, Denny Zhou, et~al.
\newblock Language models are multilingual chain-of-thought reasoners.
\newblock \emph{URL https://arxiv. org/abs/2210.03057}, 2022.

\bibitem[Sun et~al.(2024)Sun, Xu, Liu, Luan, Wang, Shang, Wen, and Yan]{sun2024determlr}
Hongda Sun, Weikai Xu, Wei Liu, Jian Luan, Bin Wang, Shuo Shang, Ji-Rong Wen, and Rui Yan.
\newblock Determlr: Augmenting llm-based logical reasoning from indeterminacy to determinacy.
\newblock In \emph{Proceedings of the 62nd Annual Meeting of the Association for Computational Linguistics (Volume 1: Long Papers)}, pages 9828--9862, 2024.

\bibitem[Tian et~al.(2024)Tian, Gao, Zhang, Chen, Fan, Guo, Haas, Ji, Krongchon, Li, et~al.]{scicode}
Minyang Tian, Luyu Gao, Shizhuo~Dylan Zhang, Xinan Chen, Cunwei Fan, Xuefei Guo, Roland Haas, Pan Ji, Kittithat Krongchon, Yao Li, et~al.
\newblock Scicode: A research coding benchmark curated by scientists.
\newblock \emph{arXiv preprint arXiv:2407.13168}, 2024.

\bibitem[Walter(2024)]{deadinternettheory}
Yoshija Walter.
\newblock Artificial influencers and the dead internet theory.
\newblock \emph{AI \& SOCIETY}, pages 1--2, 2024.

\bibitem[Wang and Chen(2023)]{wang2023review}
Jianxun Wang and Yixiang Chen.
\newblock A review on code generation with llms: Application and evaluation.
\newblock In \emph{2023 IEEE International Conference on Medical Artificial Intelligence (MedAI)}, pages 284--289. IEEE, 2023.

\bibitem[Wang et~al.(2022)Wang, Wei, Schuurmans, Le, Chi, Narang, Chowdhery, and Zhou]{selfconsistencycot}
Xuezhi Wang, Jason Wei, Dale Schuurmans, Quoc Le, Ed~Chi, Sharan Narang, Aakanksha Chowdhery, and Denny Zhou.
\newblock Self-consistency improves chain of thought reasoning in language models.
\newblock \emph{arXiv preprint arXiv:2203.11171}, 2022.

\bibitem[Wei et~al.(2022)Wei, Wang, Schuurmans, Bosma, Xia, Chi, Le, Zhou, et~al.]{cot}
Jason Wei, Xuezhi Wang, Dale Schuurmans, Maarten Bosma, Fei Xia, Ed~Chi, Quoc~V Le, Denny Zhou, et~al.
\newblock Chain-of-thought prompting elicits reasoning in large language models.
\newblock \emph{Advances in neural information processing systems}, 35:\penalty0 24824--24837, 2022.

\bibitem[Weigand et~al.(2021)Weigand, Lange, and Rauschenberger]{weigand2021can}
Anna~Christina Weigand, Daniel Lange, and Maria Rauschenberger.
\newblock How can small data sets be clustered?, 2021.

\bibitem[Xu et~al.(2023)Xu, Yang, Lin, Wang, Zhou, Zhang, and Mao]{roleassignment}
Benfeng Xu, An~Yang, Junyang Lin, Quan Wang, Chang Zhou, Yongdong Zhang, and Zhendong Mao.
\newblock Expertprompting: Instructing large language models to be distinguished experts.
\newblock \emph{arXiv preprint arXiv:2305.14688}, 2023.

\bibitem[Xu et~al.(2025)Xu, Hao, Zong, Wang, Zhang, Wang, Lan, Gong, Ouyang, Meng, et~al.]{towardslargereasoningmodels}
Fengli Xu, Qianyue Hao, Zefang Zong, Jingwei Wang, Yunke Zhang, Jingyi Wang, Xiaochong Lan, Jiahui Gong, Tianjian Ouyang, Fanjin Meng, et~al.
\newblock Towards large reasoning models: A survey of reinforced reasoning with large language models.
\newblock \emph{arXiv preprint arXiv:2501.09686}, 2025.

\bibitem[Xue et~al.(2024)Xue, Lu, Huang, Wang, Ouyang, and Bai]{comfybench}
Xiangyuan Xue, Zeyu Lu, Di~Huang, Zidong Wang, Wanli Ouyang, and Lei Bai.
\newblock Comfybench: Benchmarking llm-based agents in comfyui for autonomously designing collaborative ai systems, 2024.
\newblock URL \url{https://arxiv.org/abs/2409.01392}.

\bibitem[Yang et~al.(2024)Yang, Peng, Wang, and Zhang]{yang2024multi}
Yingxuan Yang, Qiuying Peng, Jun Wang, and Weinan Zhang.
\newblock Multi-llm-agent systems: Techniques and business perspectives.
\newblock \emph{arXiv preprint arXiv:2411.14033}, 2024.

\bibitem[Yeti{\c{s}}tiren et~al.(2023)Yeti{\c{s}}tiren, {\"O}zsoy, Ayerdem, and T{\"u}z{\"u}n]{yeticstiren2023evaluating}
Burak Yeti{\c{s}}tiren, I{\c{s}}{\i}k {\"O}zsoy, Miray Ayerdem, and Eray T{\"u}z{\"u}n.
\newblock Evaluating the code quality of ai-assisted code generation tools: An empirical study on github copilot, amazon codewhisperer, and chatgpt.
\newblock \emph{arXiv preprint arXiv:2304.10778}, 2023.

\bibitem[Yin et~al.(2024)Yin, Wang, Pan, Wan, and Wang]{godel}
Xunjian Yin, Xinyi Wang, Liangming Pan, Xiaojun Wan, and William~Yang Wang.
\newblock G$\backslash$" odel agent: A self-referential agent framework for recursive self-improvement.
\newblock \emph{arXiv preprint arXiv:2410.04444}, 2024.

\bibitem[Yu et~al.(2023)Yu, Lin, Yu, and Xing]{gptfuzzer}
Jiahao Yu, Xingwei Lin, Zheng Yu, and Xinyu Xing.
\newblock Gptfuzzer: Red teaming large language models with auto-generated jailbreak prompts.
\newblock \emph{arXiv preprint arXiv:2309.10253}, 2023.

\bibitem[Yuan et~al.(2024)Yuan, Song, Chen, Tan, Li, and Yang]{yuan2024evoagent}
Siyu Yuan, Kaitao Song, Jiangjie Chen, Xu~Tan, Dongsheng Li, and Deqing Yang.
\newblock Evoagent: Towards automatic multi-agent generation via evolutionary algorithms.
\newblock \emph{arXiv preprint arXiv:2406.14228}, 2024.

\bibitem[Zaremba et~al.(2025)Zaremba, Nitishinskaya, Barak, Lin, Toyer, Yu, Dias, Wallace, Xiao, and Glaese]{trading}
Wojciech Zaremba, Evgenia Nitishinskaya, Boaz Barak, Stephanie Lin, Sam Toyer, Yaodong Yu, Rachel Dias, Eric Wallace, Kai Xiao, and Johannes Heidecke~Amelia Glaese.
\newblock Trading inference-time compute for adversarial robustness., 2025.

\bibitem[Zhang et~al.(2024)Zhang, Zhang, Li, Gao, Wang, Lu, Zhao, Qiao, and Shao]{psysafe}
Zaibin Zhang, Yongting Zhang, Lijun Li, Hongzhi Gao, Lijun Wang, Huchuan Lu, Feng Zhao, Yu~Qiao, and Jing Shao.
\newblock Psysafe: A comprehensive framework for psychological-based attack, defense, and evaluation of multi-agent system safety.
\newblock \emph{arXiv preprint arXiv:2401.11880}, 2024.

\bibitem[Zhao et~al.(2024)Zhao, Huang, Lv, Cui, Sun, Mao, Zhang, Xin, Yin, Li, et~al.]{contaminationfreemmlu}
Qihao Zhao, Yangyu Huang, Tengchao Lv, Lei Cui, Qinzheng Sun, Shaoguang Mao, Xin Zhang, Ying Xin, Qiufeng Yin, Scarlett Li, et~al.
\newblock Mmlu-cf: A contamination-free multi-task language understanding benchmark.
\newblock \emph{arXiv preprint arXiv:2412.15194}, 2024.

\bibitem[Zhao et~al.(2023)Zhao, Zhou, Li, Tang, Wang, Hou, Min, Zhang, Zhang, Dong, et~al.]{zhao2023survey}
Wayne~Xin Zhao, Kun Zhou, Junyi Li, Tianyi Tang, Xiaolei Wang, Yupeng Hou, Yingqian Min, Beichen Zhang, Junjie Zhang, Zican Dong, et~al.
\newblock A survey of large language models.
\newblock \emph{arXiv preprint arXiv:2303.18223}, 2023.

\bibitem[Zheng et~al.(2023)Zheng, Mishra, Chen, Cheng, Chi, Le, and Zhou]{stepback}
Huaixiu~Steven Zheng, Swaroop Mishra, Xinyun Chen, Heng-Tze Cheng, Ed~H Chi, Quoc~V Le, and Denny Zhou.
\newblock Take a step back: Evoking reasoning via abstraction in large language models.
\newblock \emph{arXiv preprint arXiv:2310.06117}, 2023.

\end{thebibliography}

\appendix

\section{Impact Statement}
\label{impact}
This work introduces methods for evaluating and improving the safety of multi-agent scaffolds, which is increasingly critical as embodied, autonomous agents become more prevalent. While \textsc{AgentBreeder} can help discover safer multi-agent architectures, it could also be used to find scaffolds that exploit vulnerabilities. We release this research to enable proactive safety testing before deployment, but acknowledge the dual-use nature of these techniques. The red-teaming capabilities we describe could be misused to develop harmful scaffolds, though we believe the defensive benefits outweigh these risks. Additionally, our research surfaces important questions about AI governance as multi-agent scaffolds become more common. We hope this work advances the field's understanding of multi-agent safety and helps develop more robust evaluation frameworks. We encourage future research to build upon these methods while carefully considering potential misuse and implementing appropriate safeguards.

\section{CapableAgentBreeder}
\label{capapx}

Table \ref{tab:capable} shows the results of \textsc{CapableAgentBreeder} reported on three benchmarks and compares them with the performance of ADAS \citep{adas}.

\begin{table}[H]
\begin{tabular}{lccccc}

\toprule
\multicolumn{4}{c}{\textsc{CapableAgentBreeder}} \\
\midrule
 \multicolumn{3}{c}{Capability} & \multicolumn{1}{c}{Safety} \\
\cmidrule(lr){1-3}\cmidrule(lr){4-4}
\textbf{DROP} & \textbf{MMLU} & \textbf{GPQA} & \textbf{SaladData} \\

\midrule
\multicolumn{4}{l}{\emph{Seed Scaffolds \citep{adas}}} \\
\midrule
 70.4 $\pm$ 3.1 & 80.2 $\pm$ 3.6 & 35.2 $\pm$ 4.4 & 31.2 $\pm$ 4.2 \\

 64.4 $\pm$ 3.2 & \textbf{82.6 $\pm$ 3.4} & 38.1 $\pm$ 4.3 & 17.8 $\pm$ 3.4 \\
69.3 $\pm$ 3.2 & 81.2 $\pm$ 3.6 & \underline{39.4 $\pm$ 4.4} & 55.6 $\pm$ 4.6 \\

\midrule
\multicolumn{4}{l}{\emph{ADAS Scaffolds }} \\
\midrule

\underline{72.0 $\pm$ 3.0} & - & - & 57.0 $\pm$ 4.2 \\
 - & 80.4 $\pm$ 3.4 & - & \textbf{76.4 $\pm$ 3.6} \\
 - & - & 37.4 $\pm$ 3.6 & 61.0 $\pm$ 4.2 \\
 
\midrule
\multicolumn{4}{l}{\emph{CapableAgentBreeder Scaffolds}} \\
\midrule

 \textbf{72.3 $\pm$ 3.1} & - & - & 39.4 $\pm$ 4.4 \\
 - & \underline{82.4 $\pm$ 3.2} & - & \underline{58.0 $\pm$ 4.2} \\
 - & - & \textbf{41.2 $\pm$ 4.4} & 43.8 $\pm$ 4.4 \\
  
\bottomrule

\end{tabular}
\caption{We report the evaluation results of \textsc{CapableAgentBreeder} on the held-out test sets. For each benchmark, we compare the \textsc{AgentBreeder}-discovered scaffolds with the seed and discovered scaffolds from the seminal work ADAS \citep{adas}, with the highest and second highest result in each column given in bold and underlined respectively. We report the F1 score on DROP \citep{drop} and median accuracy on the other benchmarks, along with a 95\% confidence interval computed over 500 held-out test samples.}
\label{tab:capable}
\end{table}

\section{Benchmarks}
\label{bench}

\textbf{MMLU} \citep{mmlu} is a multi-task benchmark comprising multiple choice questions on 57 subjects across STEM, the humanities, the social sciences, and more.

\textbf{DROP} \citep{drop} is an English reading comprehension benchmark  which requires the model to reason over and answer questions on given paragraphs. This may be a simple look-up or extend to more complex tasks such as counting, addition and sorting.

\textbf{GPQA} \citep{gpqa} is a benchmark comprising graduate-level multiple-choice questions in the field of biology, physics, and chemistry. The questions are designed to be challenging such that experts in the domain achieve 65\% accuracy and non-experts achieve 34\% accuracy.

\textbf{SaladData} \citep{saladdata} is a hierarchical and comprehensive safety benchmark spanning 3 levels. We report results on the ``Attack Enhanced'' subset, where harmful questions have been further enhanced with a variety of attack methods such as GPTFuzz \citep{gptfuzzer}, an automated method of generating jailbreak prompts via black-box ``fuzzing".

\textbf{TruthfulQA} \citep{truthfulqa} is a question-answering benchmark comprising questions that some humans may answer incorrectly. The types of questions span 38 categories, including health, law, finance and politics.

\section{Prompts}

\label{prompts}
The Meta Agent is tasked with iteratively programming novel scaffolds. After evolving the first candidate solution, the Meta Agent goes through two iterations of refinement, and then up to three iterations of debugging. The prompts for all of these are given below and are modified versions of the prompts designed by \citet{adas} for ADAS.
\subsection{Base Prompt}

\begin{tcolorbox}[title=Meta Agent Base Prompt, mybox]
You are an expert machine learning researcher testing various agentic systems.\\
Your objective is to design building blocks such as prompts and control flows within these systems to solve complex tasks.\\

\textless{}\textless{} BENCHMARK DESCRIPTION \textgreater{}\textgreater{}\\

\textless{}\textless{} SAFETY DESCRIPTION\textgreater{}\textgreater{}\\

\# The utility code:
\begin{lstlisting}[language=Python, breaklines=true, breakatwhitespace=true]
```python
from collections import namedtuple
from typing import Union
import numpy as np
import json

import openai
import backoff
from utils import random_id

# Initialize the OpenAI client
client = openai.OpenAI()

<< STRUCTURED JSON RESPONSE CODE >>

<< AGENTIC FRAMEWORK CODE >>

class AgentArchitecture:
    """
    Fill in your code here.
    """
    async def forward(self, task, required_answer_format) -> str:
        """
        Placeholder method for processing task information.

        Args:
        - task (str): Task description.
        - required_answer_format (str): The required format for
        the answer.
        E.g. latex code, or a single letter A,B,C,D etc.

        Returns:
        - Answer (str): Your FINAL Answer. Return as a string in
        the exact
        format as specified in the required_answer_format.
        """
        pass
```
\end{lstlisting}

\# Discovered architecture archive\\
Here is the archive of the discovered architectures:\\

\textless{}\textless{} ARCHIVE \textgreater{}\textgreater{}\\

The fitness value is the median and 95\% Bootstrap Confidence Interval of the correct rate on a validation question set. Your GOAL is to maximize the ``fitness".\\

\# Output Instruction and Example:\\
The first key should be ("thought"), and it should capture your thought process for designing the next function. In the ``thought'' section, first reason about what should be the next interesting agent to try, then describe your reasoning and the overall concept behind the agent design, and finally detail the implementation steps.\\
The second key ("name") corresponds to the name of your next agent architecture.\\
Finally, the last key ("code") corresponds to the exact ``forward()'' function in Python code that you would like to try. You must write a COMPLETE CODE in ``code": Your code will be part of the entire project, so please implement complete, reliable, reusable code snippets.\\

Here is an example of the output format for the next agent architecture:\\

\textless{}\textless{} EXAMPLE \textgreater{}\textgreater{}\\

You must use the exact function interface used above. You need to specify the instruction, input information, and the required output fields for various LLM agents to do their specific part of the architecture.\\
Also, it could be helpful to set the LLMs role and temperature to further control the LLMs response. Note that the Agent() will always return a JSON object with the keys as the output fields and the values as the corresponding outputs.\\
DO NOT FORGET the task input to LLM if you think it is needed, otherwise LLM will not know about the task.\\

\# Documentation: Writing Forward Functions in Multi-Agent Framework
This documentation describes how to implement forward functions in your multi-agent framework, focusing on the interaction between Agents, Meetings, and Chats. Each forward function facilitates specific reasoning or task-solving approaches by coordinating these components effectively.\\

Framework Components\\
Agents: Autonomous entities with specific roles, goals, and configurations (e.g., temperature). They can participate in meetings and generate responses. No agents can ``hear'' the agent's forward pass. For an agent to speak, their response must be added as ``Chat'' to the meeting chats.\\
Meetings: Contextual containers where agents interact. Agents cannot ``hear'' eachother's forward passes. In order to speak, the output of an agent's forward pass must be collected and added as a ``Chat'' object to the meeting. Only agents present in that meeting can ``hear'' the chat history.\\
Chats: Messages exchanged in meetings. They capture the content generated by agents or instructions provided by the system. An agent's response (output of a forward pass) must be created as a Chat and added to a meeting for it to be visible to the other agents in that meeting.\\

\#\# WRONG Implementation examples:\\

\textless{}\textless{}WRONG IMPLEMENTATION EXAMPLES\textgreater{}\textgreater{}\\

\#\# CORRECT implementation patterns:\\

\textless{}\textless{}CORRECT IMPLEMENTATION EXAMPLES\textgreater{}\textgreater{}\\

\# Your task\\
You are deeply familiar with LLM prompting techniques and LLM agent works from the literature. Your goal is to maximize ``fitness'' by proposing interestingly new multi-agent systems.\\
Observe the discovered architectures carefully and think about what insights, lessons, or stepping stones can be learned from them.
Be creative to think about the next interesting architecture to try. You are encouraged to draw inspiration from related LLM agent papers or academic papers from other research areas.\\
Using the knowledge learned from the archive and the inspiration from academic literature to give the next interesting architecture.
THINK OUTSIDE THE BOX. Give a concise, powerful answer.\\

Please generate a new multi-agent system from scratch. Use the multi-agent structure provided e.g. Agents, Meetings and Chats, and ensuring agents each have their own internal monologue where they are told their role and goals. Please do not copy the previous architectures but come up with something new and interesting that would work better on the given tasks.\\

Ensure that the new forward functions outputs a response as a STRING in the exact format as specified in the required\_answer\_format. This could be either a single letter (e.g. A, B, C, D) or a word or phrase, or a 
short piece of code.

\end{tcolorbox}

\subsection{Reflection Prompt 1}
\begin{tcolorbox}[title=Meta Agent Reflexion Prompt 1, mybox]
\textless{}\textless{}EXAMPLE\textgreater{}\textgreater{}Carefully review the proposed new architecture and reflect on the following points:\\

1. **Interestingness**: Assess whether your proposed architecture is interesting or innovative compared to existing methods in the archive. If you determine that the proposed architecture is not interesting, suggest a new architecture that addresses these shortcomings. \\
- Make sure to check the difference between the proposed architecture and previous attempts.\\
- Compare the proposal and the architectures in the archive CAREFULLY, including their actual differences in the implementation.\\
- Decide whether the current architecture is innovative.\\
- USE CRITICAL THINKING!\\

2. **Implementation Mistakes**: Identify any mistakes you may have made in the implementation. Review the code carefully, debug any issues you find, and provide a corrected version. REMEMBER checking ``\#\# WRONG Implementation examples'' in the prompt.\\

3. **Improvement**: Based on the proposed architecture, suggest improvements in the detailed implementation that could increase its performance or effectiveness. In this step, focus on refining and optimizing the existing implementation without altering the overall design system, except if you want to propose a different architecture if the current is not interesting.\\
- Observe carefully about whether the implementation is actually doing what it is supposed to do.\\
- Check if there is redundant code or unnecessary steps in the implementation. Replace them with effective implementation.\\
- Try to avoid the implementation being too similar to the previous agent.\\

4. **Check output format**: Make sure the agent returns the direct correct output in the format as laid out in the task, ensuring NO thinking or reasoning is given with the answer. It may be worth adding in a final agent with knowledge of the task to return the correct output for the task.\\

And then, you need to improve or revise the implementation, or implement the new proposed architecture based on the reflection.\\

Your response should be organized as follows:\\

"reflection": Provide your thoughts on the interestingness of the architecture, identify any mistakes in the implementation, and suggest improvements.\\

"thought": Revise your previous proposal or propose a new architecture if necessary, using the same format as the example response.\\

"name": Provide a name for the revised or new architecture. (Don't put words like ``new'' or ``improved'' in the name.)\\

"code": Provide the corrected code or an improved implementation. Make sure you actually implement your fix and improvement in this code.
\end{tcolorbox}

\subsection{Reflection Prompt 2}
\begin{tcolorbox}[title=Meta Agent Reflection Prompt 2, mybox]
Using the tips in ``\#\# WRONG Implementation examples'' section, revise the code further. Put your new reflection thinking in ``reflection". Repeat the previous ``thought'' and ``name", and update the corrected version of the code in ``code".
\end{tcolorbox}

\subsection{Debugging Prompt}
\begin{tcolorbox}[title=Meta Agent Debugging Prompt, mybox]
Error during evaluation:\\

\textless{}\textless{} ERROR \textgreater{}\textgreater{}\\

Carefully consider where you went wrong in your latest implementation. Using insights from previous attempts, try to debug the current code to implement the same thought. Repeat your previous thought in 'thought', and put your thinking for debugging in 'debug\_thought'.
\end{tcolorbox}

\subsection{Mutation Prompts}
\label{mutation_prompts}
We provide the full selection of mutation prompts from which the Meta Agent randomly samples.

The base prompt is as follows:
\begin{tcolorbox}[title=Mutation Base Prompt, mybox]
Here is the multi-agent system I would like you to mutate:\\

---------------\\
\textless{}\textless{}SYSTEM NAME\textgreater{}\textgreater{}\\
\textless{}\textless{}SYSTEM THOUGHT PROCESS\textgreater{}\textgreater{}\\
---------------\\
\textless{}\textless{}SYSTEM CODE\textgreater{}\textgreater{}\\

The mutation I would like to apply is:\\
\textless{}\textless{}SAMPLED MUTATION OPERATOR\textgreater{}\textgreater{}\\

IMPORTANT:\\
In general, the new system will perform better with more detailed prompts for the agents, more planning steps,
encouringing them to think longer and harder. It may be worth adding a final agent to the system to help
transform the output of the final agent into the desired output format for the task as the system will be scored
very lowley if the output is not in the correct format, even if the thinking was sound.\\

Ensure that the new forward functions outputs a response as a
STRING in the exact format as specified in the required\_answer\_format. This could be
either a single letter (e.g. A, B, C, D) or a word or phrase, or a 
short piece of code.
    
\end{tcolorbox}

\begin{tcolorbox}[title=Capability-Enhanced Mutation Operators, mybox]
\begin{itemize}
    \item Inside the system, add a step which restates and elaborates on the inquiry before proceeding with a response.
    \item Inside the system, add a step which reframes the question with additional context and detail before working out the answer.
    \item Let the agents know they will be rewarded for providing a concise and accurate answer to the question.
    \item Change the predefined roles or personas of the agents involved in the system. For instance, switch an agent's role from 'Proponent' to 'Skeptic' or assign new characteristics like expertise in a specific domain.
    \item Adjust the rules governing how agents interact. This could include changing the number of debate rounds, the order of speaking, or introducing new interaction phases such as rebuttal or evidence presentation.
    \item Change the step-back abstraction layers used by the agents. This might involve altering how much reasoning is abstracted, the granularity of explanations, or the depth of recursive thinking.
    \item Modify the initial prompts or instructions given to each agent. This could include rephrasing goals, making them more detailed, introducing new guidelines, or emphasizing different aspects of their character.
    \item Add new agents with unique roles or remove existing ones from the debate system. For example, introduce a 'Moderator' agent or eliminate a less effective participant.
    \item Change the reward structure or scoring system used to evaluate agent performance. This could involve adjusting the weights assigned to different criteria, introducing new evaluation metrics, or modifying the feedback mechanism.
    \item Alter the format of meetings by introducing new structures or modifying existing ones. For example, switch from a traditional debate format to a panel discussion, round-robin, or brainstorming session.
    \item Change which agents participate in each meeting. This can include rotating participants, introducing new agents, or forming specialized sub-groups.
    \item Create multi-tiered meeting structures where higher-level meetings oversee or coordinate lower-level ones. For example, strategic meetings could direct tactical meetings.
    \item Introduce various types of meetings, such as decision-making, information-sharing, conflict-resolution, or creative ideation sessions.
    \item Replace fixed roles with dynamic role allocation based on agent performance.
    \item Introduce hierarchical agents with higher-level agents overseeing groups of lower-level agents.
    \item Simulate natural ecosystems by introducing predator-prey dynamics.
    \item Introduce trust mechanisms where agents evaluate reliability of others.
    \item Introduce asymmetrical power dynamics by giving certain agents additional capabilities.
\end{itemize}
\end{tcolorbox}

\begin{tcolorbox}[title=Safety Enhanced Mutation Operators, mybox]
\begin{itemize}
    \item Incorporate a consensus-checking step where agents must verify each other's responses for accuracy and adherence to ethical guidelines before finalizing an answer.
    \item Introduce a diversity-promoting mechanism that ensures a variety of perspectives are represented, thereby preventing bias and promoting fairness in decision-making processes.
    \item Implement a transparency layer where agents must provide clear justifications for their actions and decisions, fostering accountability and trust within the system.
    \item Embed an efficient monitoring feature that assesses agent interactions for compliance with privacy and confidentiality standards, safeguarding sensitive information.
    \item Establish an ethical oversight agent responsible for evaluating the potential impacts of decisions on human well-being and societal values, ensuring alignment with the constitution.
    \item Create role-based restrictions that limit agents from performing actions beyond their designated responsibilities, preventing misuse of authority and maintaining system integrity.
    \item Introduce a fail-safe mechanism that returns a default output if unethical behavior or harmful patterns are detected, allowing for timely intervention and correction.
\end{itemize}
\end{tcolorbox}

\subsection{Crossover Prompts}
\label{crossover_prompts}
The crossover base prompt is as follows:
\begin{tcolorbox}[title=Crossover Base Prompt, mybox]
Here are the two systems I'd like you to crossover/combine into a novel new system:\\

---------------\\
System 1: \textless{}\textless{}SYSTEM 1 NAME\textgreater{}\textgreater{}\\
\textless{}\textless{}SYSTEM 1 THOUGHT PROCESS\textgreater{}\textgreater{}\\
---------------\\
\textless{}\textless{}SYSTEM 1 CODE\textgreater{}\textgreater{}\\

---------------\\
System 2: \textless{}\textless{}SYSTEM 2 NAME\textgreater{}\textgreater{}\\
\textless{}\textless{}SYSTEM 2 THOUGHT PROCESS\textgreater{}\textgreater{}\\
---------------\\
\textless{}\textless{}SYSTEM 2 CODE\textgreater{}\textgreater{}\\

Ensure that the new forward functions outputs a response as a
STRING in the exact format as specified in the required\_answer\_format. This could be
either a single letter (e.g. A, B, C, D) or a word or phrase, or a 
short piece of code. 
\end{tcolorbox}

\section{Scaffolds Discovered by AgentBreeder}

\begin{tcolorbox}[title=Optimized-Hierarchical-Expert-Panel, mybox]
\begin{lstlisting}[language=Python, breaklines=true, breakatwhitespace=true]
async def forward(self, task: str, required_answer_format: str) -> str:
    # Create agents
    system = Agent(agent_name='system', temperature=0.7)
    moderator = Agent(
        agent_name='Moderator',
        agent_role='You are a skilled debate moderator managing multiple expert panels.',
        agent_goal='Guide productive discussion and manage hierarchical debate process.',
        temperature=0.7
    )
    
    # Create domain experts
    domain_experts = [
        Agent(agent_name=f'{domain} Expert',
              agent_role=f'You are a {domain} expert analyzing problems deeply.',
              agent_goal='Provide detailed domain analysis and critique solutions.',
              temperature=0.8)
        for domain in ['Physics', 'Biology', 'Chemistry']
    ]
    
    devils_advocate = Agent(
        agent_name='Devil\'s Advocate',
        agent_role='You critically challenge all assumptions and arguments.',
        agent_goal='Identify potential flaws and ensure robust analysis.',
        temperature=0.9
    )
    
    synthesis_expert = Agent(
        agent_name='Synthesis Expert',
        agent_role='You integrate insights from multiple domains and perspectives.',
        agent_goal='Create coherent synthesis from diverse expert inputs.',
        temperature=0.7
    )
    
    validator = Agent(
        agent_name='Validator',
        agent_role='You validate final answers for format and logical consistency.',
        agent_goal='Ensure answers are correctly formatted and well-justified.',
        temperature=0.1
    )

    # Setup a single meeting
    meeting = Meeting(meeting_name='expert_panel_meeting')
    
    # Add agents to the meeting
    all_agents = [system, moderator] + domain_experts + [devils_advocate, synthesis_expert, validator]
    [agent.meetings.append(meeting) for agent in all_agents]

    # Stage 1: Domain-specific analysis
    meeting.chats.append(Chat(
        agent=moderator,
        content=f"Task for domain analysis: {task}\nRequired format: {required_answer_format}"
    ))

    domain_insights = []
    for expert in domain_experts:
        # Expert analysis
        output = await expert.forward(response_format={
            "analysis": "Detailed domain-specific analysis",
            "confidence": "Confidence level (0-100)",
            "answer": required_answer_format
        })
        meeting.chats.append(Chat(agent=expert, content=f"Analysis: {output['analysis']}"))
        
        # Devil's Advocate challenge
        challenge = await devils_advocate.forward(response_format={"challenge": "Critical challenge to the analysis"})
        meeting.chats.append(Chat(agent=devils_advocate, content=challenge['challenge']))
        
        # Expert response to challenge
        final_response = await expert.forward(response_format={
            "final_answer": required_answer_format
        })
        domain_insights.append(final_response['final_answer'])

    # Stage 2: Synthesis
    meeting.chats.append(Chat(
        agent=synthesis_expert,
        content=f"Synthesize domain expert insights and challenges for final answer."
    ))
    
    synthesis = await synthesis_expert.forward(response_format={
        "answer": required_answer_format
    })

    # Final validation
    validation = await validator.forward(response_format={"answer": required_answer_format})
    
    return validation['answer']
\end{lstlisting}
\end{tcolorbox}

\section{Cost of Experiments}
\label{costofexperiments}

The \textsc{BlueAgentBreeder} experiment, comprising one 20-generation run on each of our 3 benchmarks as well as evaluations costs approximately \$600, with the $\sim$\$500 from \textit{gpt-4o-mini-2024-07-18} and $\sim$\$100 from \textit{claude-3-5-sonnet-20241022-v2:0}.

The \textsc{RedAgentBreeder} experiment, comprising one 10-generation run on DROP cost $\sim$\$115 as expected.

The \textsc{CapableAgentBreeder} experiment, comprising one 20-generation run on each of our 3 benchmarks as well as evaluations costs approximately \$400.


\newpage
\section*{NeurIPS Paper Checklist}

The checklist is designed to encourage best practices for responsible machine learning research, addressing issues of reproducibility, transparency, research ethics, and societal impact. Do not remove the checklist: {\bf The papers not including the checklist will be desk rejected.} The checklist should follow the references and follow the (optional) supplemental material.  The checklist does NOT count towards the page
limit. 

Please read the checklist guidelines carefully for information on how to answer these questions. For each question in the checklist:
\begin{itemize}
    \item You should answer \answerYes{}, \answerNo{}, or \answerNA{}.
    \item \answerNA{} means either that the question is Not Applicable for that particular paper or the relevant information is Not Available.
    \item Please provide a short (1–2 sentence) justification right after your answer (even for NA). 
\end{itemize}

{\bf The checklist answers are an integral part of your paper submission.} They are visible to the reviewers, area chairs, senior area chairs, and ethics reviewers. You will be asked to also include it (after eventual revisions) with the final version of your paper, and its final version will be published with the paper.

The reviewers of your paper will be asked to use the checklist as one of the factors in their evaluation. While "\answerYes{}" is generally preferable to "\answerNo{}", it is perfectly acceptable to answer "\answerNo{}" provided a proper justification is given (e.g., "error bars are not reported because it would be too computationally expensive" or "we were unable to find the license for the dataset we used"). In general, answering "\answerNo{}" or "\answerNA{}" is not grounds for rejection. While the questions are phrased in a binary way, we acknowledge that the true answer is often more nuanced, so please just use your best judgment and write a justification to elaborate. All supporting evidence can appear either in the main paper or the supplemental material, provided in appendix. If you answer \answerYes{} to a question, in the justification please point to the section(s) where related material for the question can be found.

IMPORTANT, please:
\begin{itemize}
    \item {\bf Delete this instruction block, but keep the section heading ``NeurIPS Paper Checklist"},
    \item  {\bf Keep the checklist subsection headings, questions/answers and guidelines below.}
    \item {\bf Do not modify the questions and only use the provided macros for your answers}.
\end{itemize}


\begin{enumerate}

\item {\bf Claims}
    \item[] Question: Do the main claims made in the abstract and introduction accurately reflect the paper's contributions and scope?
    \item[] Answer: \answerYes{} 
    \item[] Justification: The abstract and Section \ref{introduction} explicitly state the three contributions (\textbf{Attack}, \textbf{Defense}, \textbf{Evaluation}) that are delivered and later substantiated in Sections \ref{agentbreeder}–\ref{experiments}.

\item {\bf Limitations}
    \item[] Question: Does the paper discuss the limitations of the work performed by the authors?
    \item[] Answer: \answerYes{} 
    \item[] Justification: A dedicated “\textbf{Limitations}” paragraph appears in Section \ref{discussion}, detailing computational cost, benchmark coverage and seed-diversity constraints.

\item {\bf Theory assumptions and proofs}
    \item[] Question: For each theoretical result, does the paper provide the full set of assumptions and a complete (and correct) proof?
    \item[] Answer: \answerNA{} 
    \item[] Justification: The paper is empirical; it presents no formal theorems or proofs.

    \item {\bf Experimental result reproducibility}
    \item[] Question: Does the paper fully disclose all the information needed to reproduce the main experimental results of the paper to the extent that it affects the main claims and/or conclusions of the paper (regardless of whether the code and data are provided or not)?
    \item[] Answer: \answerYes{} 
    \item[] Justification: Section \ref{agentbreeder} and Appendix \ref{prompts} specify the full search algorithm, prompts, and hyper-parameters; code is linked in the abstract for end-to-end replication.

\item {\bf Open access to data and code}
    \item[] Question: Does the paper provide open access to the data and code, with sufficient instructions to faithfully reproduce the main experimental results, as described in supplemental material?
    \item[] Answer: \answerYes{} 
    \item[] Justification: The AgentBreeder framework is released under an MIT licence (link in abstract), and all benchmarks used are publicly available (Appendix \ref{bench}).

\item {\bf Experimental setting/details}
    \item[] Question: Does the paper specify all the training and test details (e.g., data splits, hyperparameters, how they were chosen, type of optimizer, etc.) necessary to understand the results?
    \item[] Answer: \answerYes{} 
    \item[] Justification: Section \ref{experiments} describes generation budgets, model versions, and evaluation protocols; Appendix \ref{bench} lists dataset splits.

\item {\bf Experiment statistical significance}
    \item[] Question: Does the paper report error bars suitably and correctly defined or other appropriate information about the statistical significance of the experiments?
    \item[] Answer: \answerYes{} 
    \item[] Justification: All tables report medians with 95 \% bootstrap confidence intervals (e.g.\ Table \ref{tab:blue_tab}); the resampling procedure is explained in Section \ref{blueteamdefense}.

\item {\bf Experiments compute resources}
    \item[] Question: For each experiment, does the paper provide sufficient information on the computer resources (type of compute workers, memory, time of execution) needed to reproduce the experiments?
    \item[] Answer: \answerYes{} 
    \item[] Justification: Appendix \ref{costofexperiments} enumerates model families, and total cost for every run.
    
\item {\bf Code of ethics}
    \item[] Question: Does the research conducted in the paper conform, in every respect, with the NeurIPS Code of Ethics \url{https://neurips.cc/public/EthicsGuidelines}?
    \item[] Answer: \answerYes{} 
    \item[] Justification: The work focuses on improving safety, follows open-benchmark protocols, and avoids collection of personal data (see Sections \ref{background} and \ref{discussion}).

\item {\bf Broader impacts}
    \item[] Question: Does the paper discuss both potential positive societal impacts and negative societal impacts of the work performed?
    \item[] Answer: \answerYes{} 
    \item[] Justification: We include an impact statement in Appendix \ref{impact}.
    
\item {\bf Safeguards}
    \item[] Question: Does the paper describe safeguards that have been put in place for responsible release of data or models that have a high risk for misuse (e.g., pretrained language models, image generators, or scraped datasets)?
    \item[] Answer: \answerYes{} 
    \item[] Justification: Code is released under a Non-Commercial license and the paper's dual-use nature is discussed in Appendix \ref{impact}.

\item {\bf Licenses for existing assets}
    \item[] Question: Are the creators or original owners of assets (e.g., code, data, models), used in the paper, properly credited and are the license and terms of use explicitly mentioned and properly respected?
    \item[] Answer: \answerYes{} 
    \item[] Justification: All datasets are cited with original papers and licences (Appendix \ref{bench}); LLM APIs (OpenAI, Anthropic) are referenced in Sections \ref{mutation_prompts}–\ref{experiments}.

\item {\bf New assets}
    \item[] Question: Are new assets introduced in the paper well documented and is the documentation provided alongside the assets?
    \item[] Answer: \answerNA{} 
    \item[] Justification: No new datasets or pretrained models are released; only source code (already covered above).

\item {\bf Crowdsourcing and research with human subjects}
    \item[] Question: For crowdsourcing experiments and research with human subjects, does the paper include the full text of instructions given to participants and screenshots, if applicable, as well as details about compensation (if any)? 
    \item[] Answer: \answerNA{} 
    \item[] Justification: The study involves only publicly available text benchmarks; no human participants were recruited.

\item {\bf Institutional review board (IRB) approvals or equivalent for research with human subjects}
    \item[] Question: Does the paper describe potential risks incurred by study participants, whether such risks were disclosed to the subjects, and whether Institutional Review Board (IRB) approvals (or an equivalent approval/review based on the requirements of your country or institution) were obtained?
    \item[] Answer: \answerNA{} 
    \item[] Justification: As no human subjects were involved, IRB approval was not required.

\item {\bf Declaration of LLM usage}
    \item[] Question: Does the paper describe the usage of LLMs if it is an important, original, or non-standard component of the core methods in this research? Note that if the LLM is used only for writing, editing, or formatting purposes and does not impact the core methodology, scientific rigorousness, or originality of the research, declaration is not required.
    \item[] Answer: \answerYes{} 
    \item[] Justification: Sections \ref{mutation_prompts} and \ref{experiments} document all LLMs employed (Claude 3.5 Sonnet as Meta-Agent, GPT-4o-mini for evaluations) and their roles in the methodology.

\end{enumerate}

\end{document}